\documentclass{article}

\usepackage{arxiv}
\usepackage[utf8]{inputenc} 
\usepackage[T1]{fontenc}    
\usepackage{hyperref}       
\usepackage{bookmark}
\usepackage{url}            
\usepackage{booktabs}       
\usepackage{amsfonts}       
\usepackage[tableposition=top]{caption}
\usepackage{graphicx}
\usepackage{amsmath, amssymb} 
\usepackage{braket}
\usepackage{float}
\usepackage[semibold,tabular]{sourceserifpro} 
\usepackage{subcaption}
\usepackage[version=4]{mhchem}
\usepackage{verbatim}
\usepackage[sorting=none]{biblatex}
\usepackage{authblk}
\usepackage{footnote}
\usepackage{siunitx}
\usepackage{xfrac}
\DeclareSIUnit\angstrom{\text {Å}}
\usepackage[frozencache=true,cachedir=minted-cache]{minted}
\usepackage{xcolor}
\definecolor{backcolour}{rgb}{0.95,0.95,0.92}

\newcommand{\gpcc}{\si{\gram\per\centi\meter\cubed}}
\newcommand{\evpa}{\si{\eV\per\angstrom}}
\newcommand{\ASE}{\texttt{ASE}~}

\title{GEARS H: Accurate machine-learned Hamiltonians for next-generation device-scale modeling}

\title{GEARS H: Accurate machine-learned Hamiltonians for next-generation device-scale modeling}

\author{Anubhab Haldar\thanks{These authors contributed equally}}
\author{Ali K. Hamze$^*$}
\author{Nikhil Sivadas}
\author{Yongwoo Shin\thanks{email: yongwoo.s@samsung.com}}
\affil{Advanced Materials Lab \\ 
Samsung Advanced Institute of Technology-America \\
Samsung Semiconductor Inc.\\
Cambridge, Massachusetts 02138, USA}

\bibliography{GEARS_ref}

\begin{document}
\maketitle

\begin{abstract}

We introduce GEARS H, a state-of-the-art machine-learned Hamiltonian framework for large-scale electronic structure simulations. 
Using GEARS H, we present a statistical analysis of the hole concentration induced in defective \ce{WSe2} interfaced with \ce{Ni}-doped amorphous \ce{HfO2} as a function of the \ce{Ni} doping rate, system density, and \ce{Se} vacancy rate in 72 systems ranging from \numrange{3326}{4160} atoms---a quantity and scale of interface electronic structure calculation beyond the reach of conventional density functional theory codes and other machine-learning-based methods.
We further demonstrate the versatility of our architecture by training models for a molecular system, 2D materials with and without defects, solid solution crystals, and bulk amorphous systems with covalent and ionic bonds.
The mean absolute error of the inferred Hamiltonian matrix elements from the validation set is below \SI{2.4}{\milli\eV} for all of these models.
GEARS H outperforms other proposed machine-learned Hamiltonian frameworks, and our results indicate that machine-learned Hamiltonian methods, starting with GEARS H, are now production-ready techniques for DFT-accuracy device-scale simulation.
    
\end{abstract}

\section{Introduction}


Density functional theory (DFT) has proven to be the most widely applied computational technique in condensed matter physics. 
Indeed, two foundational DFT papers rank among the top 10 most-cited papers of all time \cite{vannoordenNTheseAreMostcited2025}. 
The applications of DFT, however, have been limited to relatively small systems due to the high computational cost of calculations. 
Systems with $\mathcal{O}(10^0-10^1)$ atoms are readily accessible, while systems with $\mathcal{O}(10^2)$ atoms require researchers to consider whether they are necessary. 
Only in recent years, with the advent of GPUs and of more powerful CPUs have system with low-$\mathcal{O}(10^3)$ atoms become possible, but such calculations are rarely done due to their exorbitant cost.

Meanwhile, progress in semiconductor device manufacturing is becoming increasingly difficult due to material and process constraints.
While ever increasing transistor densities were once taken for granted, now, other solutions must be sought. 
These include new materials like 2D transition metal dichalcogenides (TMDs) as channel materials and new device geometries like monolithic 3D integrated circuits \cite{dhananjayITCSIMonolithic3DIntegrated2021, zengNREETransistorEngineeringBased2024, palNEThreedimensionalTransistorsTwodimensional2024}. 
Exploration of new materials and fabrication of devices with novel transistor geometries, however, requires large upfront investment. 
This presents an opportunity for new, low-cost computational methods to lead industry forward. 
Such new methods, ideally, will not sacrifice the accuracy of DFT in pursuit of device-scale simulation.

In this work, we present GEARS Hamiltonian (GEARS H), our framework for machine-learned Hamiltonians (MLH) in a linear combination of atomic orbitals (LCAO) basis. 
GEARS H is the first MLH framework to enable models of realistic, device-scale systems that are beyond the reach of traditional DFT, demonstrating the true strength of MLH methods. 
We show this by training a model on a combined system of $\ce{Ni}$-doped amorphous $\ce{HfO2}$ interfaced to $\ce{WSe2}$. 
This system was recently proposed \cite{sivadasModulationDopingControl2025} for modulation doping of $\ce{WSe2}$. 
We use the model to perform a statistical study of the hole concentration induced in the \ce{WSe2} layer in device-scale systems (\numrange{3326}{4160} atoms) as a function of $\ce{Ni}$ doping rate, \ce{Se} vacancy rate, and system density.


We further demonstrate the broad applicability of GEARS H by applying it to 1) Lithium Bis(trifluoromethanesulfonyl)imide (\ce{LiTFSI}), a molecular system with 6 elements, 2) 2D $\ce{WSe_{2-x}}$ $\left(0.0 \leq  x < 0.07\right)$, 3) a dataset of nine different 2D materials featuring eight atomic species, 4) $\ce{Ag_x Au_{1-x}}$ $\left(0.34 < x < 0.72\right)$, a metal alloy, 5) amorphous \ce{SiO2} (a-\ce{SiO2}), a covalent solid, and 5) amorphous \ce{HfO2} (a-\ce{HfO2}), a mixed ionic-covalent solid.

Our results suggest that, with the advancements presented in our framework, MLH models are now production-ready tools for next-generation device modeling.
There has been a dramatic acceleration in the search for new crystalline structures through the successful development and deployment of machine-learning-based interatomic potentials (MLIPs) \cite{merchantNScalingDeepLearning2023,kaplanFoundationalPotentialEnergy2025, yangMatterSimDeepLearning2024}.
We hope that GEARS H leads to a similar phenomenon in the field of electronic structure.

GEARS H builds on previous work towards MLHs.
The earliest attempts include those by Hegde and Bowen \cite{hegdeSRMachinelearnedApproximationsDensity2017} and Schutt \textit{et al}. \cite{schuttNCUnifyingMachineLearning2019}.
Advances were made by Li and colleagues \cite{liNCSDeeplearningDensityFunctional2022} and the related work by Gong \textit{et al} \cite{gongNCGeneralFrameworkEquivariant2023}.
Unke and colleagues \cite{unkeSE3equivariantPredictionMolecular2021} have demonstrated highly accurate learning of molecular Hamiltonians and provide mathematical details for the construction of such models.
Nigam and colleagues \cite{nigamJCPEquivariantRepresentationsMolecular2022} demonstrate linear models of molecular Hamiltonians with rigorous mathematical analysis.
Several \verb|e3nn|-based \cite{geigerE3nnEuclideanNeural2022} models have also been developed including DeepH-E3 \cite{gongNCGeneralFrameworkEquivariant2023}, DeePTB \cite{guNCDeepLearningTightbinding2024}, QHNet \cite{yuEfficientEquivariantGraph2023}, and HamGNN \cite{zhongnCMTransferableEquivariantGraph2023}.
Only one other MLH framework by Xia \textit{et al}. has been applied to amorphous systems \cite{xiaLearningElectronicHamiltonian2025}.
To the best of our knowledge, GEARS H has the fewest number of parameters of any MLH model reported in the literature (12\%, 8.3\%, and 3\% the parameter count of DeeTB-E3\cite{guNCDeepLearningTightbinding2024}, DeepH-E3\cite{gongNCGeneralFrameworkEquivariant2023}, and HamGnn\cite{zhongnCMTransferableEquivariantGraph2023}, respectively), and is the only model that has been successfully applied to amorphous systems interfaced with other geometries.

The Hamiltonian architecture of GEARS H is inspired by architectural decisions in PhiSNet, ACEhamiltonians, and DeepH-E3.
We present the ideas underlying GEARS H, provide a user- and performance-focused implementation of our model named \verb|gears_h|, made using E3x \cite{unkeE3xMathrmEquivariant2024}.
We also provide a companion data processing package named \verb|gears_h_tools|, which supplies an interface to GPAW \cite{mortensenJCPGPAWOpenPython2024} (which we choose because it is open-source, written in Python, easy to install, and allows for low-cost training data generation using strictly confined numerical atomic orbitals and projector-augmented waves to describe core electrons \cite{larsenPRBLocalizedAtomicBasis2009}).
Interfaces to other LCAO codes are possible and we welcome community contributions to implement data conversion to the format required for GEARS H.
The packages are available on Github and are linked at the end of the paper.
Our work is a part of a greater ongoing effort which we call GEARS (\underline{\bf G}iant-scale \underline{\bf E}lectronic structure and \underline{\bf A}tomic configuration \underline{\bf R}esearch \underline{\bf S}olution) that will be further detailed in subsequent publications.

\section{Results and Discussion}

\begin{figure}[h]
    \centering
    \includegraphics[width=1.0\linewidth]{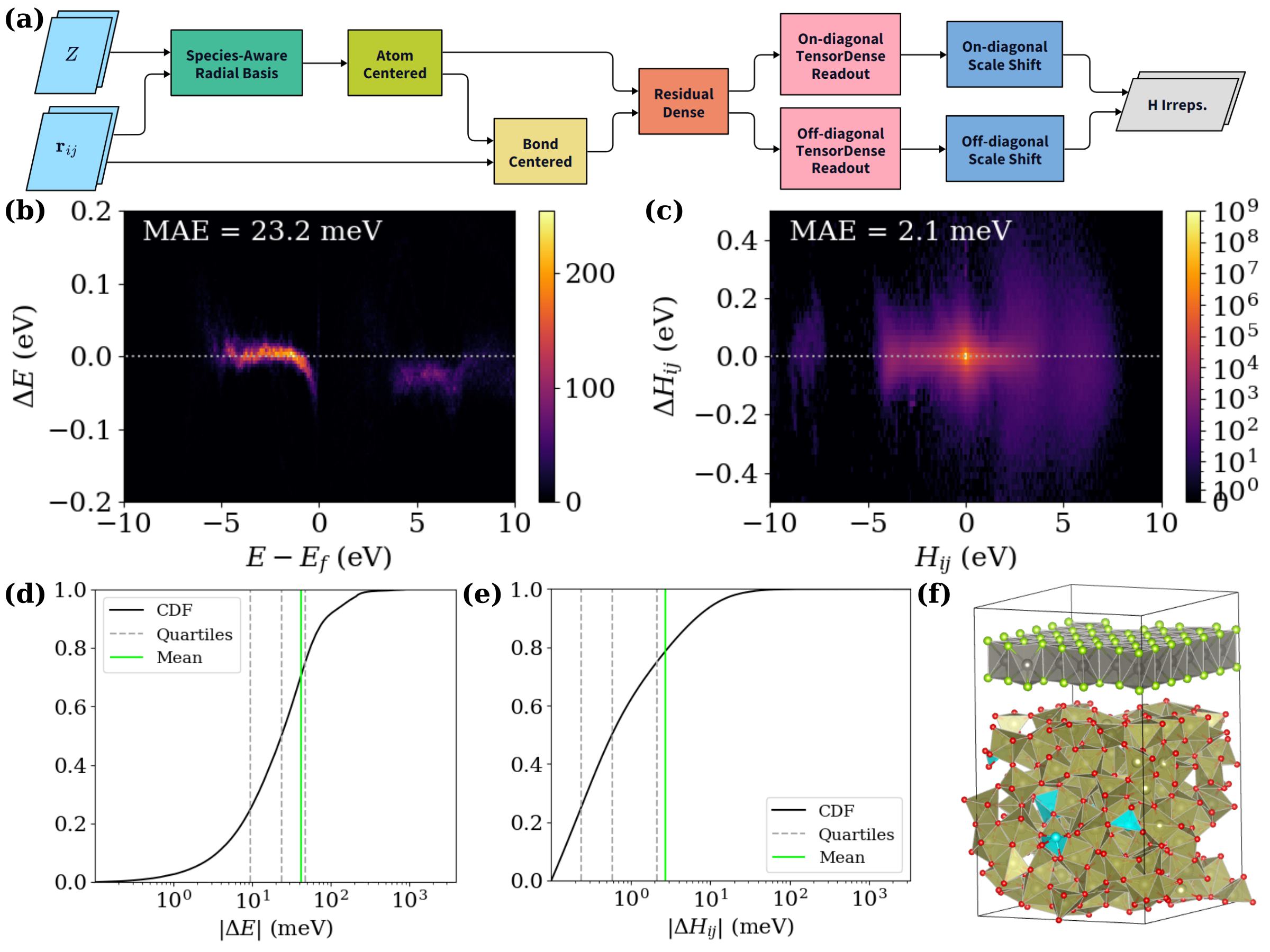}
    \caption{(a) GEARS H architecture overview. (b) Validation set eigenvalue errors relative to the reference eigenvalues. (c) Validation set Hamiltonian matrix element errors relative to the reference matrix elements with an \texttt{asinh} scale to help resolve bins that have smaller counts. (d) Cumulative distribution function of the eigenvalue errors larger than \SI{0.1}{\milli\eV}. (e) Cumulative distribution function of the Hamiltonian matrix element errors larger than \SI{0.1}{\milli\eV}. (f) Sample structure from the training dataset. \ce{Hf} are gold, \ce{Ni} are turquoise, \ce{O} are red, \ce{W} are gray, and \ce{Se} are green. The MAEs shown in (b-c) are averaged across training set structures, whereas the mean values in (d-e) are taken across all errors. GEARS H performs very well on this highly complex system.}
    \label{fig:hero_fig}
\end{figure}

\subsection{Model architecture}

An overview of the GEARS H architecture is shown in Fig. \ref{fig:hero_fig}(a). 
Here, we briefly describe the model inputs and outputs and present the details of three pieces of the model architecture: the atom-centered descriptor, the bond-centered descriptor, and the scale-shift layers. 
Detailed layer architecture diagrams, descriptions of the other layers \cite{he2ICCVPRCDeepResidualLearning2016}, additional discussion of the layers described here, and a brief discussion of the choices made in developing the model can be found in the Supplementary Information.

The input data consists of an array of atomic numbers $Z_i$, sparse neighbor lists, and the corresponding pairwise Cartesian vectors. 
The training and output data consists of two arrays of Hamiltonian irreducible representations (irreps) in direct-sum form corresponding to atom-centered and atom-pair interaction Hamiltonian blocks similar to the approach used in \cite{unkeSE3equivariantPredictionMolecular2021}.

\subsubsection{The atom-centered descriptor}

Atom-centered descriptors have been extensively studied in the context of MLIPs; we refer the reader to the review by Musil {\it et al.} \cite{musilCRPhysicsInspiredStructuralRepresentations2021} on their design choices. 
The inputs to the atom-centered descriptor in GEARS H are expansions of the local neighborhoods of atoms using a 2-body (2B) basis consisting of radial and angular functions. 

To make our descriptor many-body and increase its sensitivity to the local atomic environment, we use the density trick\cite{shapeevMMSMomentTensorPotentials2016, drautzPRBAtomicClusterExpansion2019, musilCRPhysicsInspiredStructuralRepresentations2021}: the outer product of {\it pooled} 2B features leads to 3-body features, and subsequent outer products lead to higher body-order features.
The descriptors created from these outer products are known as atom-centered density correlations (ACDC)\cite{nigamJCPUnifiedTheoryAtomcentered2022}.
We focus on learning a dense subspace since higher-order descriptors in the Atomic Cluster Expansion (ACE) model are known to be relatively sparse\cite{darbynCMCompressingLocalAtomic2022}.
We do this using a \verb|TensorDense| layer as implemented in \texttt{E3x} \cite{unkeE3xMathrmEquivariant2024} to learn a feature-wise tensor product of two linear projections of the input features. 
2B descriptors are passed through \verb|TensorDense| layers (optionally, although we recommend at least one--otherwise, the descriptor does not have many-body information) to get 3B descriptors, and so on.
The pooling operation is a sum over the atoms $j$ in the neighborhood of a given atom $i$, which we do using the \verb|indexed_sum| operation implemented in \verb|E3x|.
Empirically, a body order of 3-5 is sufficient for acceptable accuracy of learned quantities like energies and forces\cite{batatiaMACEHigherOrder2023}---we find the same is true for GEARS H.

An essential difference between the well-explored previous energy predictions and our approach is that our model does {\it not} average over all rotations of these ACDCs. 

These 2B, 3B, ..., $(2N - 1)$B descriptors are then separately (and optionally) message-passed between atoms using self-attention (SA) to carry out the learnable coupling across all incoming messages.
In this work, none of the models presented make use of SA, so we leave discussion of it to the Supplementary Information.
Crucially, the omission of message-passing does not reduce our accuracy and simultaneously greatly reduces our parameter count, contributing to the status of GEARS H as the smallest reported MLH model in the literature.
In the Supplementary Information, we present a comparison of one of the models shown below against identical models with 1 and 2 message-passing steps, which perform worse than the model with no message-passing steps.

The separate (optionally) message-passed descriptors are then sent through a nonlinear block, for which we use a two-layer perceptron with a residual connection. The dense layers are interleaved with a \verb|LayerNorm| and \verb|mish| activation function to refine the atom-centered features and add functional expressivity.

Finally, the resulting descriptors are reduced to a user-controlled maximum angular momentum and then concatenated along the feature dimension ($F$ in \verb|E3x| convention).
By keeping the descriptors of distinct body-order separate until the very end, the intervening message-passing and nonlinear blocks remain small (block diagonal in body order), which further helps reduce parameter counts and speed up training.

\subsubsection{The bond-centered descriptor}

Off-diagonal terms in Hamiltonian or overlap matrix blocks can be predicted as a function of atom-centered features of the two atoms comprising a `bond'. Here, a bond refers to any two atoms with significant basis function overlap (and corresponding interaction strength).
To calculate atom-pairwise features for predicting off-diagonal matrix blocks, we sum pairs of atom-centered features, which is similar to the approach used in PhiSNet \cite{unkeSE3equivariantPredictionMolecular2021}. 
To add more functional expressivity, the pooled features are refined using a two-layer perceptron with a \verb|LayerNorm| and \verb|mish| activation between layers and a residual connection between layers, akin to the nonlinear block in the atom-centered descriptor.
For bond orientation information, we expand the bond vector into radial and angular basis functions, which we pass through a \verb|Dense| layer as a learnable linear projection to refine features.
Finally, we  take a feature-wise tensor product of linearly-projected bond vector expansion with the pooled atom-pair features. 

\subsubsection{The scale-shift layers}

The scale-shift layers are non-learnable blocks that scale and shift the parity-symmetric scalars in the readout output. 
This allows (scalar) outputs from the readout to be approximately zero-centered and unit-variance by mapping the outputs to the physical values, which can vary greatly in magnitude.
These parameters can be extracted from the training dataset, a functionality which we have built into GEARS H.

\subsection{Case study: Modulation doping of \ce{WSe2} with \ce{Ni}-doped a-\ce{HfO2}}

Transition metal dichalogenides (TMDs) like \ce{WSe2} are attractive candidates for post-silicon channel materials because they are atomically thin and can have mobilities comparable to \ce{Si}.
Conventional substitution doping strategies of TMDs, however, lead to reduced mobilities due to the introduction of scattering centers, and do not contribute enough carriers to the TMD.
Recently, Sivadas and Shin \cite{sivadasModulationDopingControl2025} proposed modulation doping of \ce{WSe2} through doping an interfaced \ce{HfO2} gate dielectric layer.
However, their work was plane-wave DFT-based, which limited the accessible doping rates and dopant distributions, prevented the consideration of the effect of \ce{Se} defects in \ce{WSe2} (which are known to form during synthesis)
, and restricted their study to crystalline \ce{HfO2} for the gate dielectric, despite the ubiquity of amorphous gate oxides in real devices.

GEARS H does not suffer from these constraints.
As a proof of concept of its utility in modeling multi-component systems of engineering interest, we train a model for amorphous, \ce{Ni}-doped \ce{HfO2} interfaced with \ce{WSe2} containing \ce{Se} vacancies.
This system presents both geometric and chemical challenges for ML-based modeling and provides a testbed for the atomistic modeling of device-scale geometries.
A large number of diverse chemical environments are present in this system, ranging from 2D crystalline \ce{WSe2} to the amorphous \ce{HfO2} bulk, which is further complicated by the presence of \ce{Ni} dopants, \ce{Se} vacancies, and the interface with \ce{WSe2}.
To our knowledge, no other MLH framework has been applied to a system of this complexity.

\subsubsection{Validation set}

A sample training structure for this system is shown in Fig. \ref{fig:hero_fig}(f).
200 structures total were generated, which were split into 160 training structures and 40 validation structures (see Methods for more details).

In Fig. \ref{fig:hero_fig}(b), we show the validation set errors of eigenvalues from inferred Hamiltonians. Within $\pm\SI{5}{\eV}$ of the Fermi level $E_f$, the eigenvalue mean absolute error (MAE) averaged across validation set systems is \SI{23.2}{\milli\eV}.
While a small increase in the error is visible at the valence band maximum, this is in fact a numerical artifact arising from uncorrelated sorting of the eigenvalues between the eigenvalues of the reference Hamiltonian and the eigenvalues of the inferred Hamiltonian.
To provide another view of the eigenvalue errors, in Fig. \ref{fig:hero_fig}(d), we show the cumulative distribution function (CDF) of the absolute eigenvalue errors larger than \SI{0.1}{\milli\eV}, and plot the quartiles and MAE of all validation set eigenvalues taken together. 
50\% of the eigenvalue errors are smaller than \SI{23.6}{\milli\eV}, which is below thermal fluctuations at room temperature $\left(k_B T|_{T=\SI{298}{\kelvin}} = \SI{25.7}{\milli\eV}\right)$.
The MAE in Fig. \ref{fig:hero_fig}(d) is higher than that shown in Fig. \ref{fig:hero_fig}(b) because, in the former, the MAE is calculated across all eigenvalues, while in the latter, we only include eigenvalues within $E_f\pm$\SI{5}{\eV}.
In other words, even when considering states more than \SI{5}{\eV} from the Fermi level, which will have correspondingly smaller impact on observables, our errors for this complex system will not impact the application of our model.

We show the Hamiltonian matrix element errors in Fig. \ref{fig:hero_fig}(c). 
Note that the color map was created using an \texttt{asinh} normalization.  
The MAE of matrix elements averaged across validation set systems is \SI{2.1}{\milli\eV}, and errors are within $\pm$\SI{1}{\eV} across the full range of matrix element values (\SIrange{-15}{55}{\eV}). 
In Fig. \ref{fig:hero_fig}(e), we show the CDF of the absolute Hamiltonian matrix element errors, after filtering out errors smaller than \SI{0.1}{\milli\eV}.
Over 90\% of errors are smaller than \SI{6}{\milli\eV}, and 50\% are smaller than \SI{0.58}{\milli\eV}.
Had all the errors been considered, the error at these quartiles would be even lower.
In the Supplementary Information, we show the on- and off-diagonal Hamiltonian matrix elements separately.
The largest errors are in on-diagonal blocks, which are also where the largest Hamiltonian matrix elements are.
This presents a clear target for future improvement, but because the on-diagonal blocks are large, the impact of the larger errors is somewhat mitigated.

Altogether, these figures indicate that while the model  performs quite well, there is room for improvement towards minimizing outliers.
Conversely, outliers have an outsized effect on the MAE.
The full CDFs reveal that the model performs very well across the validation set, and can therefore be trusted for studying modulation doping of \ce{WSe2} interfaced with a \ce{Ni}-doped a-\ce{HfO2} gate dielectric.
A systematic study of the effect of random outliers and random errors in general on the eigenvalues of matrices will be critical for enhancing trust in MLH model predictions as their reliability and usage grows.

\subsubsection{Application to device-scale structures}

\begin{figure}[h]
    \centering
    \includegraphics[width=1.0\linewidth]{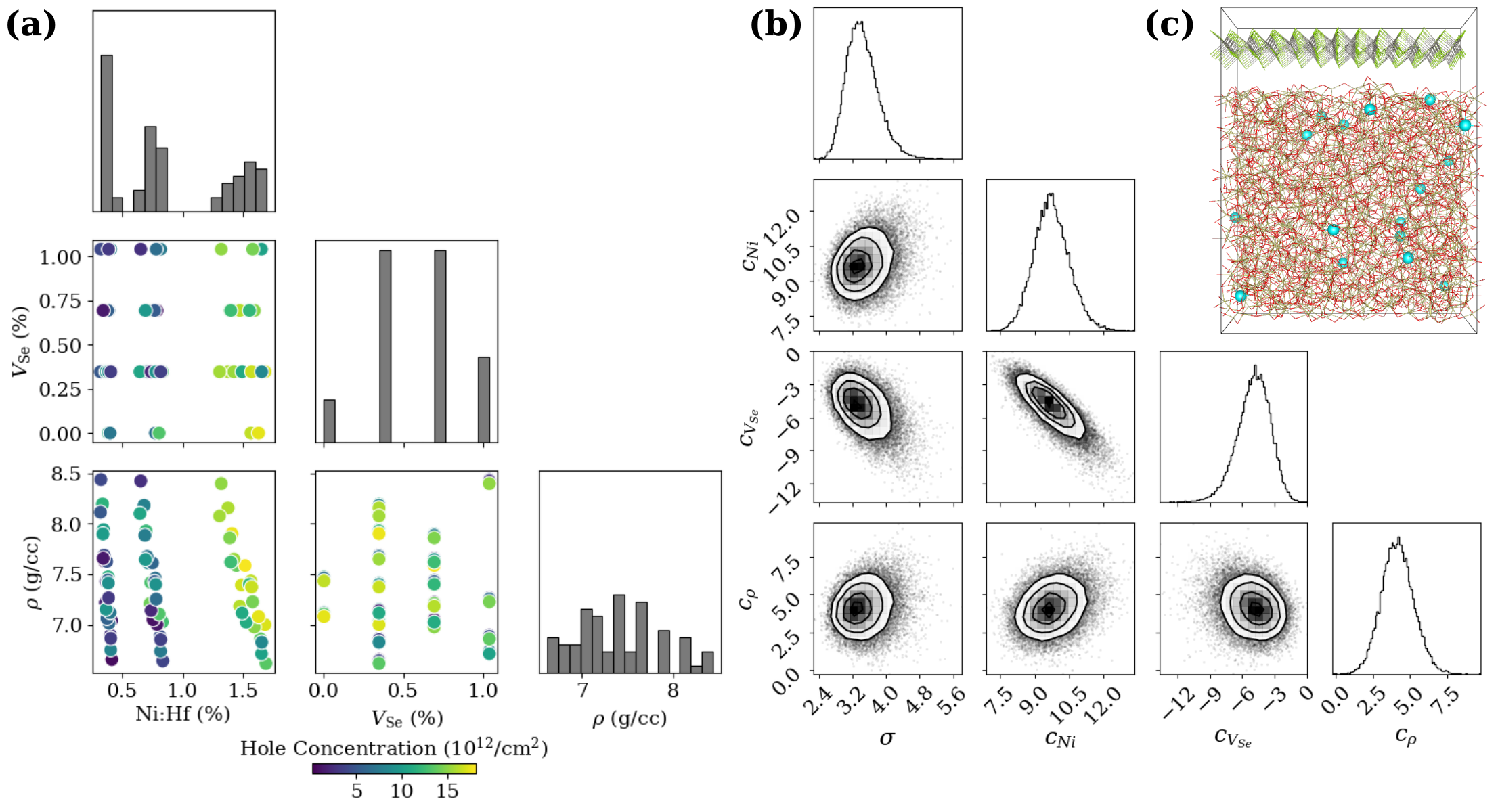}
    \caption{(a) \ce{Ni} doping rates, \ce{Se} vacancy rates, and total system density of the large scale systems used in the statistical analysis. Diagonal figures are histograms with 20 bins of the distribution of each individual quantity, while the off-diagonal figures show pair plots with the corresponding hole concentrations. (b) Results of the Bayesian analysis showing interactions between the posterior likelihood of the parameters of the model. Diagonal histograms show the distribution of each parameter and the residual $\sigma$. Off-diagonal subplots show marginal joint distributions of the parameters with iso-likelihood contours. Most interactions are weak with the exception of $c_{\rm Ni}$ and $c_{V_{\rm{Se}}}$, indicating that both the $p$-doping due to \ce{Ni} and the $n$-doping due to \ce{Se}-vacancies can be strong or weak together. (c) Selected system used in the statistical study. \ce{Hf}, \ce{O}, \ce{W}, and \ce{Se} atoms have been hidden to highlight the \ce{Ni} dopants spread through the a-\ce{HfO2}.}
    \label{fig:stats-fig}
\end{figure}

Given the good performance of our model across the validation set, we now use it to perform a statistical study of hole concentrations in the \ce{WSe2} layer. 
We considered systems sizes ranging from \numrange{3326}{4160} atoms with systems with side lengths ranging from \SIrange{3.4}{4.5}{\nano\meter}.
These systems are comparable in size to candidate next-generation 2D field effect transistors under active research \cite{wuNRDefectsFactorLimiting2016, illarionovNEUltrathinCalciumFluoride2019}.
72 structures were generated for the statistical study with \ce{Ni} doping rates ranging from $\ce{Ni}:\ce{Hf} = \numrange{3.23e-3}{16.86e-3}$, \ce{Se} vacancy rates ranging from approximately 0\%-1\% (approximately 0-$26e12$ vacancies/\si{\centi\meter\squared}), and system densities ranging from \SIrange{6.6}{8.4}{\gpcc}.
The distribution of \ce{Ni} doping rates, \ce{Se} vacancy rates, and system densities considered are shown in the histograms in the diagonal subplots of Fig. \ref{fig:stats-fig}(a), along pair plots colored with the corresponding hole concentrations in the off-diagonal subplots.

We emphasize that the full process of generating this data (the generation of all the structures, the inference of all their electronic structures, and the diagonalization of the inferred Hamiltonians) took less than 12 hours on a single GPU workstation with 8 Nvidia L40S GPUs.
The inference of the Hamiltonians itself was the fastest part of the process took approximately \SI{13}{\second} per structure (see additional discussion on inference in the Supplementary Information).
Investigations of realistic systems of this complexity and length scale would not be feasible without GEARS H.

With the \ce{WSe2} layer hole concentration data from large-scale systems as our target variable, we now perform a Bayesian study to find the effect of several experimentally-controllable parameters.
We report the parameters as $A_{B}^{C}$, where $A$ is the mean value of the parameter, $B$ is $3\%$ high-density interval, and $C$ is $97\%$ high-density interval.  

We \textit{ansatz} a simple linear model dependent on relevant, controllable design variables: the \ce{Ni}:\ce{Hf} doping rate ($\rho_{\rm Ni}$), the total system density ($\rho$), and the \ce{Se} vacancy rate ($\rho_{\rm V_{Se}}$). Concretely,
\begin{align}
    \rho_{h} = c_{\rm Ni} \rho_{\rm Ni} + c_{\rm \rho} \rho + c_{\rm V_{Se}} \rho_{\rm V_{Se}},
\end{align}
where we have shifted the densities such that they are centered at approximately \SI{0}{\gpcc}.

The results of the Bayesian analysis is shown in Fig. \ref{fig:stats-fig}(b).
The diagonal subplots are distributions of each parameter in the model and the residual, and the off-diagonal subplots are correlations between the parameters and residual themselves. 

First, we consider $c_{\rm Ni}$, the proportionality coefficient between the magnitude of hole doping of \ce{WSe2} and the \ce{Ni} doping rate.
In the histogram in the second row of Fig. \ref{fig:stats-fig}(b), we see $c_{\mathrm{Ni}}$ has a positive mean of approximately $9.7_{8.3}^{11.2}$.
This implies a strong positive correlation between the \ce{Ni} doping rate and the hole concentration in the interfaced \ce{WSe2}, since the posterior likelihood of the doping coefficient is entirely positive.
Importantly, to the best of our knowledge, this is the first time variations in induced hole concentrations due to modulation doping have been accounted for at an atomistic level. 
Our studies provide strong statistical evidence of robust \ce{Ni}-induced $p$-doping in interfaced \ce{WSe2} and greatly extend previous work \cite{sivadasModulationDopingControl2025} to amorphous gate oxides and realistic doping rates. 

Next, we focus on the effect of system density on the hole concentration in the \ce{WSe2} layer, which is represented by $c_{\rm \rho}$. 
Since the distribution of $c_{\rm \rho}$ is peaked at approximately $4.1_{2.0}^{6.2}$, and the distribution is almost entirely positive, this is strong evidence that greater system densities facilitate higher $p$-doping of the \ce{WSe2} layer.
This is strong evidence for the intuitive picture that lower densities lead to larger structural variations that can create trap states and increase the potential barrier through which the \ce{Ni} electrons tunnel through.
Both of these effects reduce the doping in the \ce{WSe2} layer.

Finally, we consider the effect of \ce{Se} vacancies on the hole doping, which is represented by $c_{\rm V_{\rm Se}}$.
The distribution of $c_{\rm V_{\rm Se}}$ is peaked at $-5.0_{-8.0}^{-2.2}$, a negative value, suggesting that \ce{Se} vacancies contribute negatively to $p$-doping in \ce{WSe2}.
In other words, there is no compensating mechanism for the $n$-doping of $\rm V_{\mathrm{Se}}$ from the gate dielectric that we find from our data.
While this relationship is weaker than the $p$-doping due to \ce{Ni}, we see that \ce{Se} vacancies and \ce{Ni} doping are competing variables in the $p$-doping of \ce{WSe2}.

We now focus on the interactions between the coefficients of the model, shown in the off-diagonal subplots in Fig. \ref{fig:stats-fig}(b). 
The interactions between the residual variable $\sigma$ and both $c_{\rm Ni}$ and $c_{\rm V_{\rm Se}}$ suggest a stronger doping effect weakly corresponds to increased residual of the model, indicating that the linear model may need additional corrections in the strong doping regime.
Very interestingly, we notice a strong correlation in the joint posterior distribution of $c_{\rm Ni}$ and $c_{\rm V_{\rm Se}}$. 
This suggests it is likely both the $p$-doping due to \ce{Ni} and $n$-doping due to \ce{Se}-vacancies can be strong or weak together, but it is very unlikely that one is strong while the other is weak.
Investigation of this correlation using more involved numerical experiments is a promising avenue for further work.
Once again, GEARS H makes this possible.

\subsection{Application to diverse chemical systems and atomic environments}

\begin{figure}[h]
    \centering
    \includegraphics[width=1.0\linewidth]{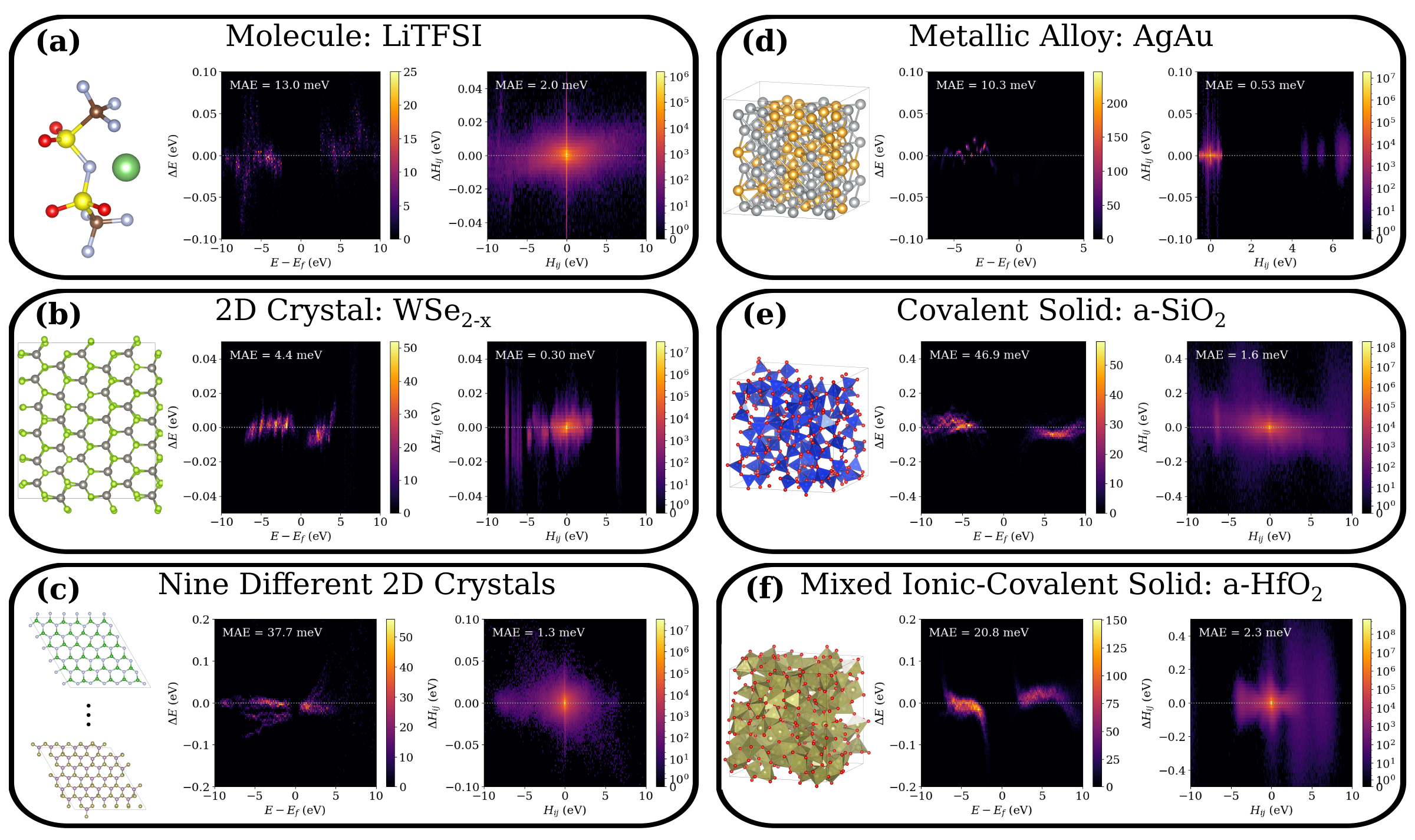}
    \caption{Sample structure from the training set and 2D histograms of eigenvalue and Hamiltonian matrix element errors of GEARS H models trained on a wide variety of materials chosen for their varied local environments. These include 1) Li-TFSI, a molecular system with 6 atomic species, 2) $\ce{WSe_{2-x}}$, a 2D TMD, 3) a collection of 9 different 2D TMDs with 8 atomic species, 4) \ce{AgAu}, a metallic alloy, 5) a-\ce{SiO2}, a covalent solid, and 6) a-\ce{HfO2}, a mixed ionic-covalent solid. Errors shown were calculated on the validation sets, and $H_{ij}$ errors are shown with an \texttt{asinh} scale. Note that error scales differ between figures, due to the variation in the range of errors. GEARS H performs extremely well across all classes of materials.}
    \label{fig:diverse_systems}
\end{figure}

In Fig. \ref{fig:diverse_systems}, we show the eigenvalue and Hamiltonian matrix element error distributions of GEARS H models trained on a diverse set of systems chosen for their widely varying atomic environments.
GEARS H can handle molecules, 2D materials, metallic alloys, amorphous solids, and systems combining these without extensive hyperparameter tuning.
Notably, in addition to the 5 species system discussed in the previous section, we include a 6 species and an 8 species system here.
Only \texttt{HamGNN} \cite{zhongnCMTransferableEquivariantGraph2023, zhongCPLUniversalMachineLearning2024} has report models with more species, and even then, they have not included defects or amorphous structures.

For each material, we show an example structure from the training dataset, the validation set eigenvalue error distribution MAE, and Hamiltonian matrix element error distribution and MAE.
Both MAEs provided are averaged MAEs across the validation set structures, and the eigenvalue MAEs are computed using eigenvalues within $\pm\SI{5}{\eV}$ of the Fermi level.
On- and off-diagonal Hamiltonian matrix element errors are presented separately in the Supplementary Information.
Additional details about the model hyperparameters along with the MAEs are shown in Table \ref{tab:example_systems}.

The first example system is \ce{LiTFSI}, which is molecular system with 6 elements.
This system presents a strong alchemical and structural challenge due to the large number of atomic species and large configurational space.
GEARS H achieves \SI{2}{\milli\eV} MAE on the Hamiltonian matrix elements.

The next set of systems we consider are 2D materials.
First, we consider $\ce{WSe_{2-x}}$.
\ce{Se} vacancies are commonly formed during synthesis, and GEARS H handles them with aplomb.
While a direct comparison is not possible with our dataset, the Hamiltonian matrix element MAE of \SI{0.30}{\milli\eV} we achieved with \ce{Se} vacancies is lower than that achieved by \texttt{DeePTB-E3}, \texttt{DeepH-E3}, and \texttt{HamGNN} on \ce{MoS2} without defects \cite{liNCSDeeplearningDensityFunctional2022, gongNCGeneralFrameworkEquivariant2023, zhongnCMTransferableEquivariantGraph2023} with 12\%, 8.3\%, and 3\% the parameter count, respectively.
In the Supplementary Information, we also show the effect of adding message-passing steps and of training set size on model accuracy for the $\ce{WSe_{2-x}}$ models.
As a combination of the many atomic species we showed in \ce{LiTFSI} and the defect \ce{WSe2}, we consider a dataset of nine defect-free binary 2D crystals comprised of eight different atomic species.
The 2D crystals included in this dataset are \ce{BN}, \ce{GeS}, \ce{GeSe}, \ce{GeTe}, \ce{MoS2}, \ce{MoSe2}, \ce{MoTe2}, \ce{WS2}, and \ce{WSe2}.
The dataset includes only 16 snapshots of each 2D crystal in the training set, and 2 of each crystal in the validation set.
Despite the wide range of atomic species and limited training data, GEARS H achieves \SI{1.3}{\milli\eV} MAE on the Hamiltonian matrix elements.

Next, we consider bulk systems.
$\ce{Ag_x Au_{1-x}}$ forms a metallic solid solution with the atoms on an FCC lattice.
Even with the wide range of compositions considered $\left(0.34 < x < 0.72\right)$, the Hamiltonian matrix element MAE is still sub-\si{\milli\eV}.

For a material with covalent bonding, we choose a-\ce{SiO2}.
This is a very challenging system to model. 
While every \ce{Si} atom is tetrahedrally coordinated by \ce{O}, there is enormous freedom in how the tetrahedra connect.
While Hamiltonian matrix element MAE is relatively low, the errors have larger variance than the other systems considered thus far.
The eigenvalues are sensitive to outlier errors in the Hamiltonian, leading to the larger MAE.
This model also performs adequately on 10 different crystalline \ce{SiO2} polymorphs (see Supplementary Information), which are not in the training or validation sets.
Better performance on this material can be achieved with optimization of the training dataset and model hyperparameters, which we did not deem necessary for this demonstration.

Finally, we consider a-\ce{HfO2}, a mixed ionic and covalent solid.
Despite the increase in the possible number of bonding environments for \ce{Hf} relative to \ce{Si} in \ce{SiO2}, the Hamiltonian matrix element MAE is still a low \SI{2.3}{\milli\eV}.

\begin{table}[h]
\centering
\begin{tabular}{@{}llllll@{}}
\toprule
System                          & Train./Val. Split & $N_{\mathrm{TD}}$ & Optimizer   & Eigenvalue MAE (meV) & $H_{ij}$ MAE (meV) \\ \midrule
\ce{LiTFSI}                     & 1200/200            & 2                 & \verb|adan| & 13.0                  & 2.0               \\
$\ce{WSe_{2-x}}$                & 160/40            & 1                 & \verb|lamb| & 4.4                  & 0.30               \\
Nine 2D                & 162/18            & 2                 & \verb|adan| & 37.7                  & 1.3               \\
$\ce{Ag_x Au_{1-x}}$                       & 96/24             & 1                 & \verb|adan| & 10.3                 & 0.53               \\
a-\ce{SiO2}                     & 140/20            & 1                 & \verb|adan| & 46.9                 & 1.6                \\
a-\ce{HfO2}                     & 160/40            & 1                 & \verb|adan| & 20.8                 & 2.3                \\ \bottomrule 
\end{tabular}
\caption{Model training and validation set sizes, number of \texttt{TensorDense} layers ($N_{\mathrm{TD}}$), the optimizer used and MAEs for the example systems shown in Fig. \ref{fig:diverse_systems}.}
\label{tab:example_systems}
\end{table}

\section{Conclusion}

We present GEARS H, a state-of-the-art MLH framework that can be applied to the widest range of chemical systems and atomic environments of any MLH framework reported in the literature.
We have made the code for our implementation of this model and for a companion data processing tool available on Github.

Using GEARS H, we train a model on a \ce{Ni}-doped a-\ce{HfO2} gate oxide interfaced with \ce{WSe2} and use the model to perform a statistical study on realistic, device-scale systems.
We analyze the effect of the \ce{Ni} doping rate, system density, and \ce{Se} vacancy rate on the induced hole concentration in the \ce{WSe2} layer.
This is a direct example of first-principles simulations, atomistic deep learning, and statistical modeling being leveraged to guide future scientific and engineering exploration for novel semiconductor design.
GEARS H makes this kind of study feasible by bypassing lengthy self-consistent cycles required by Kohn-Sham DFT---the inference of the large-scale structures takes just $\sim \SI{13}{\second}$ on our hardware.

We further demonstrate the remarkable flexibility of GEARS H by training models on molecular, 2D, metallic alloy, amorphous covalent solids, and mixed ionic-covalent solid systems. 
In all cases, the MAE of the Hamiltonian matrix elements is smaller than \SI{2.4}{\milli\eV}.

With GEARS H, MLH frameworks are now production-ready tools for next-generation device modeling.

\section{Methods}

\subsection{Training and validation data generation}

\subsubsection{Structure generation}

To generate the defect \ce{WSe2} structures, We use the \verb|mx2| build module in \ASE \cite{hjorthlarsenJPCMAtomicSimulationEnvironment2017} with a $6 \times 3$ orthogonal supercell of \ce{WSe2}. The primitive cell is generated with a lattice constant of 3.32 \si{\angstrom}, and a thickness (vertical spacing between \ce{Se} atoms) of 3.2 \si{\angstrom}, with 2.3 \si{\angstrom} of vacuum (for subsequent stacking of \ce{Ni}-doped \ce{HfO2}). The precise values the lattice constants and thickness are not optimized, since the structures are annealed and relaxed later, and the underlying variation in strain is an intended variability in the dataset.
A uniform random diagonal strain of $\pm 2 \%$ is applied to the lattice constants.
A random number of \ce{Se} vacancies is incorporate. A Poisson distributed with a mean of 1.0 is used.

The a-\ce{HfO2} training snapshots were generated using \texttt{Packmol}\cite{martinezJCCPACKMOLPackageBuilding2009} and \ce{HfO2} ``molecules". Target densities were randomly chosen from a uniform random distribution between 7.0 $\pm$ 1.0 \gpcc. The \ce{HfO2} geometries were then optimized using the LBFGS optimizer in \ASE to a maximum force of 0.2 \evpa, followed by a piecewise-constant-temperature anneal from 800 \si{\kelvin} to 400 \si{\kelvin} in steps of -100 \si{\kelvin}, running for 2000 steps to 400 steps, in steps of -400 steps. The anneals were done using the Bussi velocity rescaling thermostat\cite{bussiJCPCanonicalSamplingVelocity2007} in the NVT ensemble, as implemented in ASE. A timestep of 2.0 \si{\femto\second} was used, with a 100 \si{\femto\second} thermostat coupling constant. The snapshots were finally optimized using LBFGS to a maximum force of 0.1 \evpa.
We found that amorphous structures generated using conventional melt-quench methods provide similar structures for amorphous \ce{HfO2}.

For \ce{Ni}-doped a-\ce{HfO2} interfaced with defect \ce{WSe2}, we start with the same initial structures as used in the defect \ce{WSe2} and a-\ce{HfO2} structures discussed above.
\ce{Hf} is substitutionally doped with \ce{Ni} with a Poisson-distributed concentration with mean of 3$\%$ of the \ce{Hf} count.
We then stack the defect \ce{WSe2} \SI{2.3}{\angstrom} above the initial \ce{Ni} doped a-\ce{HfO2}.
We performed molecular dynamics with a piecewise constant annealing schedule using the Bussi thermostat \cite{bussiJCPCanonicalSamplingVelocity2007} in \ASE with a timestep of \SI{2.0}{\femto\second} and a thermostat coupling constant of $\tau = \SI{100}{\femto\second}$, using the MACE MPA-0 foundation potential \cite{batatiaMACEHigherOrder2023,batatiaDesignSpaceE3Equivariant2022, batatiaFoundationModelAtomistic2024}.
Geometries were first optimized to 0.2 \si{\eV\per\angstrom} using the LBFGS optimizer in ASE.
They were then annealed down from 800 \si{\kelvin} to 400 \si{\kelvin} in steps of -100 \si{\kelvin}, running for 2000 steps to 400 steps, in steps of -400 steps.
A final optimization using LBFGS down to a maximum force of 0.1 \si{\eV\per\angstrom} was performed.

To generate the large scale structures for the statistical study, we generate structures in the same manner as we generated the training and validation set above.

\ce{LiTFSI} training snapshots were generated starting from a single conformer of TFSI, replacing the \ce{H} with \ce{Li}. The geometries were optimized to 0.05 \evpa, followed a 2 \si{\pico\second} molecular dynamics at 100 \si{\kelvin}. The \texttt{Bussi} thermostat in \texttt{ASE} was used, with a couple time constant of 50 \si{\femto\second}. Both the optimization and molecular dynamics were performed using The MACE-MPA-0 \cite{batatiaFoundationModelAtomistic2024} foundation potential including DFT-D3 dispersion correction \cite{grimmeJCPConsistentAccurateInitio2010, takamotoNCUniversalNeuralNetwork2022}. 
The training/validation split was 1200/200 structures, owing to the small amount of data per snapshot for a single molecule.

The nine 2D system dataset includes 18 structures each of \ce{BN}, \ce{GeS}, \ce{GeSe}, \ce{GeTe}, \ce{MoS2}, \ce{MoSe2}, \ce{MoTe2}, \ce{WS2}, and \ce{WSe2}. These were generated by downloading relevant structure files from C2DB \cite{haastrup2MComputational2DMaterials2018, gjerding2MRecentProgressComputational2021} and rattling them with \texttt{ASE}.

The $\ce{Ag_x Au_{1-x}} \left(0.34 < x < 0.72\right)$ training snapshots were generated by first generating bulk silver $3\times3\times3$ supercells and then replacing $\rm N$ atoms of silver with gold, where $N \sim Poisson(\lambda=N_{atoms} / 2)$. The \ce{AgAu} geometries were then optimized using the LBFGS optimizer in \ASE to a maximum force of \SI{0.5}{\eV\per\angstrom}, including cell, but maintaining cell shape. This was followed by a piecewise-constant-temperature anneal from 1600 \si{\kelvin} to 700 \si{\kelvin} using the Bussi velocity rescaling thermostat in the NVT ensemble, as implemented in ASE. An adaptive timestep was used based on temperature-dependent heuristic to make sure atoms almost never move beyond a given distance (0.08 \si{\angstrom} per time step). A 100 \si{\femto\second} thermostant coupling constant was used. The snapshots were finally optimized using LBFGS to a maximum force of 0.1 \evpa.

The a-\ce{SiO2} dataset snapshots were generated by randomly scattering \ce{SiO2} trimers (to ensure the local stoichiometry was correct) using \texttt{Packmol}\cite{martinezJCCPACKMOLPackageBuilding2009}.
The densities of the generated structures ranged  from \SIrange{2.025}{2.4}{\gpcc}
The structures were then pre-relaxed until $F_{\mathrm{max}} \leq \SI{5}{\evpa}$.
Next, the structures were annealed from \SIrange{2300}{1000}{\kelvin} in steps of \SI{100}{\kelvin} for \SI{3}{\pico\second} per step.
Finally, the structure was relaxed until $F_{\mathrm{max}} \leq \SI{1e-2}{\evpa}$.
Both relaxations used LBFGS and the MACE-MP-0a large model, while the annealing used the MACE-MP-0a medium model.
160 structures were generated in total.

\subsubsection{LCAO DFT calculations}
\label{sec:dft}

The LCAO DFT GPAW calculations were done using the $szp$ basis sets included with GPAW. At the time of writing, GEARS generates training data using Hamiltonians at the $\Gamma$-point, so  after the electronic structures were converged, a non-self-consistent calculation was done using the converged density to extract the Hamiltonian and $S$-matrix at the $\Gamma$-point only. 
We use the generalized gradient approximation for the exchange-correlation functional \cite{perdewPRLGeneralizedGradientApproximation1996}, grid spacing of $h = 0.2$ for all datasets except the \ce{WSe2} dataset, where we used $h = 0.25$. 
Calculations were converged to a maximum change in the electron density smaller than 0.001 electrons per valence electron.
To reduce the extent of the basis functions and thereby reduce the size of the training structures and number of neighbors in the training data, the \ce{Ag}, \ce{Au}, and \ce{Hf} basis functions were confined until the atomic eigenstates shifted up by \SI{0.3}{\eV}.
This is a strong confinement, but we expect minimal effects due to the large number of basis functions available in the bulk.
For the \ce{Ni} atoms, we place the $3p$ electrons in the core and use an effective on-site interaction interaction of $U_{\mathrm{eff}} = \SI{4.5}{\eV}$ on the $3d$ electrons.
For \ce{Ag}, we froze the $4p$ electrons in the core.

The GEARS H GPAW interface in \verb|gears_h_tools|, our companion data processing package for \verb|gears_h|, currently requires (due to the GPAW API used) that each atom must have unique neighbors---that is, each atom cannot have an interaction with an atom and the periodic images of the same atom.
(We aim to remove this restriction and add interfaces to additional LCAO codes in the future.)
This sets a minimum cell size of $2\times$ the longest basis function in the system.
For \ce{WSe2}, \ce{HfO2}, and the combined \ce{HfO2}:\ce{Ni} + \ce{WSe2} datasets, the longest cutoff length was \SI{8.0}{\angstrom}.
For \ce{SiO2}, since we did not confine the \ce{Si} basis functions, the longest cutoff length was \SI{8.4}{\angstrom}.
With the confinement of the \ce{Ag} and \ce{Au} basis functions, the cutoff for the \ce{AgAu} dataset was \SI{6.1}{\angstrom}.
Finally, for the many 2D dataset, we used a cutoff length of \SI{7.5}{\angstrom}.

\subsection{Data splitting}

All datasets were randomly shuffled before being split into training and validation sets. 
For the \ce{SiO2} dataset, where we ensured an equal representation of densities in the training and validation sets.
The data was split before training to prevent data leakage.

\subsection{Hole concentration calculation}

Using the GEARS H model detailed in Fig. \ref{fig:hero_fig}, we infer the Hamiltonians of the 72 large-scale structures generated as discussed above.
To get the eigenvalues, we require an $S$-matrix, which we compute for each structure.
The $S$-matrix is computed pairwise across atoms and is therefore not computationally intensive to generate.
Using the $S$-matrix and inferred Hamiltonian, we solve the generalized eigenvalue problem to get the eigenvalues, which we then shift such that the Fermi level is at \SI{0}{\eV}.
We then species-project the density of states (DOS) and then smooth the projected DOSes using \num{20,000} points with Gaussians of width \SI{0.05}{\eV}.

To get the hole concentration from the projected, smoothed DOSes, we integrate the \ce{W}- and \ce{Se}-projected DOS from the Fermi level to \SI{0.2}{\eV} above the Fermi level.

\subsection{Bayesian analysis}

We consider weakly regularizing priors for the coefficients and residual.
For $c_{\rm Ni}$, a normal distribution centered at 15 with a standard deviation of 10 is used as a priorl; for $c_{\rm V_{\rm Se}}$, a normal distribution centered at -7 with a standard deviation of 10; for $c_{\rm \rho}$, a normal distribution centered at 5 with a standard deviation of 10. For the residual $\sigma$, a half-Cauchy distribution with $\beta=10$ is used.
The model is sampled over 8 chains, each for 4000 samples, with 2000 samples of burn-in using \verb|PyMC| v5.23.0 \cite{abril-plaPCSPyMCModernComprehensive2023} with the default No-U-Turn sampler.

\subsection{Machine-learned H model}

There is very little variation in the model hyperparameters used to train the models presented in this work.
A full configuration (broken into its distinct pieces) is provided in Section \ref{s:S-hyperparams}. Here, we only briefly discuss the most important hyperparameters used and which were varied between models.

In the \texttt{data} section, \verb|n_train|, \verb|n_valid|, \verb|atoms_pad_multiple|, and \verb|nl_pad_multiple| are changed depending on the dataset.
The first two control the number of training and validation structures, while the last two control the number of recompilations of the model (through the maximum amount of padding a neighbor list array and an atomic species array is permitted) that the user is willing to allow.
The total number of training and validation structures used for each model is shown in Table \ref{tab:example_systems}.

The only change made in the \texttt{atom\_centered} section across models is the number of \verb|TensorDense|s, which was set to 1 or 2 for all models shown in this work.
The radial basis is an input to the atom-centered descriptor and is included as a subsection of \texttt{atom\_centered}.
The only change made between models in \texttt{radial\_basis} is to adjust the cutoff radius to the maximum basis set cutoff across species in each dataset.
Basis set cutoffs are discussed in Section \ref{sec:dft}.

Similarly, for the \texttt{bond\_centered} section, all hyperparameters were left unchanged except for the cutoff, which was set to the largest cutoff across species in the dataset (see Section \ref{sec:dft}).

The residual dense layer (controlled by the \texttt{mlp} section of the config) was left unchanged across all models trained in this work.
Three layers were used with output feature sizes of 32, 16, and 32, in that order.
We use a \texttt{bent\_identity} nonlinear activation function between layers.

The only changes made in the \texttt{optimizer} section was to switch between the \texttt{adan} \cite{xieAdanAdaptiveNesterov2024} and \texttt{lamb} \cite{youLargeBatchOptimization2020} optimizers, depending on which resulted in a lower loss model.
\texttt{adan} was best for all models except the defect \ce{WSe2} model.
Which optimizer was used for each model is shown in \ref{tab:example_systems}.
The only learning rate schedule changes made were to adjust the \texttt{accumulation\_size} parameter to make the \texttt{reduce\_on\_plateau} scheduler check if the loss had plateaued only once per epoch.

\subsubsection{Loss function}

We use a combination of mean-squared error (MSE) and root-mean-squared error (RMSE) as the default loss function.
This loss is invariant to rotations of vectors, whereas the mean absolute error is not \cite{wangDesignSpaceMolecular2024}. 
RMSE loss functions have a nonzero gradient at the first order, even in the neighborhood of 0 difference between ground truth data and predicted output.
The weight between the MSE and RMSE losses is controllable by the user, as is the weight between on- and off-diagonal losses.

For this work, we weighted the MSE and RMSE evenly, and we weighted the off-diagonal block irreps 4 times more than the on-diagonal blocks.
Weighting the off-diagonal irreps higher than the on-diagonal irreps helped mitigate overfitting of the on-diagonal blocks, which was more likely to occur due to the dearth of on-diagonal block training data relative to off-diagonal block training data.
Loss parameters were left unchanged across all models.



\section{Code, Dataset, and Model Availability}

We have made available two packages.
The first, \texttt{gears\_h} (available at \url{https://github.com/SamsungDS/gears\_h} ), implements the model architecture presented in this paper and is used for training new models and using existing models for inference.
The second package, \texttt{gears\_h\_tools} (available at \url{https://github.com/SamsungDS/gears\_h\_tools} ), processes training data from LCAO codes to the format needed for training \texttt{gears\_h}.

The raw training data and inference data is too large to make available, but we share the training and validation structures at \url{https://doi.org/10.5281/zenodo.17808475}, from which new \texttt{GPAW} training data can be generated.
We also share the trained model checkpoints at \url{https://doi.org/10.5281/zenodo.17808323}, which can be used for inference.

\printbibliography

\clearpage

%
%
\appendix
\newrefsection

\renewcommand{\appendixname}{Supplementary Information}

\renewcommand{\thepage}{S\arabic{page}}
\renewcommand{\thesection}{S\arabic{section}}
\renewcommand{\thetable}{S\arabic{table}}
\renewcommand{\thefigure}{S\arabic{figure}}
\renewcommand{\figurename}{Supplementary Figure}
\setcounter{figure}{0}
\setcounter{table}{0}
\setcounter{section}{1}

\section*{Supplementary Information}
\setcounter{figure}{0}
\setcounter{table}{0}
\setcounter{page}{1}
\setcounter{section}{0}

\section{Machine-learned Hamiltonian}


Kohn-Sham density functional theory (KS-DFT) involves solving the following generalized eigenvalue problem:
\
\begin{align}
    \mathbf{H \psi} = E \mathbf{S \psi}
\end{align}

in a self-consistent manner. The Hamiltonian operator $\mathbf H$ describes all electronic interactions. In KS-DFT, $\mathbf H$ is constructed within an independent-particle {\it ansatz}.
In a strictly localized numerical orbital basis where the basis functions vanish completely outside a cutoff radius, the equation above appears as follows:
\
\begin{align}
    \mathbf{H_{MM'} C_{M'n}} = \mathbf{\epsilon_n S_{MM'} C_{M'n}}
\end{align}

Here, $\mathbf{M,M'}$ are basis function indices, $\mathbf{n}$ is the eigenstate index, $\mathbf{H}$ is the Hamiltonian, $\mathbf{S}$ is the overlap matrix between basis functions, and $\mathbf{C}$ is the set of eigenvectors of the Hamiltonian in the localized numerical atomic orbital basis. The blocks of $\mathbf H$  corresponding to the interaction between two atoms are identically zero outside of cutoff radii, motivating the approximation of these matrix blocks using recent methods of atomistic machine learning using localized atom-centered density correlations and message-passing.  

Solving the eigenvalue problem can be done using a linear combination of atomic orbitals (LCAO) approach, which strongly motivates further research in the surrogate modeling of the Hamiltonian operator. In particular, a sufficiently good approximation of the self-consistent Hamiltonian has the following benefits:

\begin{itemize}
    \item The self-consistency loop is avoided, eliminating the most time-consuming step in KS-DFT.
    \item The calculation of matrix elements is avoided, reducing the computational load.
\end{itemize}



It is essential to organize the design principles to explain our surrogate Hamiltonian for investigating large-scale amorphous systems.
1) The architecture should be enable tradeoffs between accuracy and resource scalability in terms of predictions for computational flexibility. 
2) Learnable parameters should be heuristically interpretable. 
3) As many physical symmetries as possible should be encoded into the architecture itself. 
This includes E(3) symmetries geometrically and permutation invariance for element-pair operations.
On top of these design principles, the architecture must be extendable and flexible. 

Our architectural choices aim to approximate the Hamiltonian blocks for atom pairs as an E(3)-equivariant and permutation-invariant function of local atomic environments.

\section{GEARS H model architecture}
\label{s:architecture}

\begin{figure}[h]
    \centering
    \includegraphics[width=1.0\linewidth]{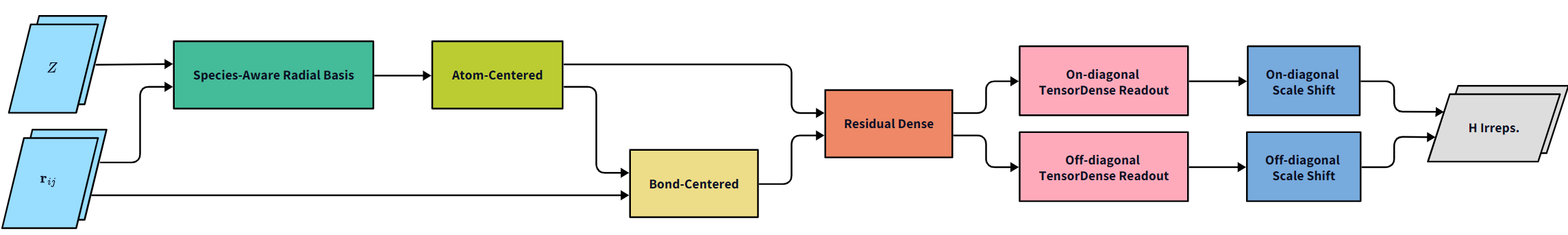}
    \caption{GEARS H architecture overview. The blocks discussed in detail below share colors with the blocks in the overview.}
    \label{fig:arch-0-overview}
\end{figure}

The input data for GEARS H consists of the atomic numbers, 2 sparse neighbor lists (one for the atom-centered environments, another for the calculation of off-diagonal Hamiltonian irreps), and the corresponding pairwise vectors. We calculate and store the "bond-centered" neighbor list for training structures during the generation of the Hamiltonian matrix blocks, but we calculate the "atom-centered" neighbor list when reading in the dataset. Reference Hamiltonian blocks for atom-centered and atom-pair interactions are converted into irreducible representations (irreps) in direct-sum form and collected in a feature-wise-irrep manner, similar to the approach in \cite{unkeSE3equivariantPredictionMolecular2021}. The output data contained two arrays of irreps corresponding to atom-centered and atom-pair interactions Hamiltonian blocks in their direct-sum form.

The GEARS H architecture (shown in Fig. \ref{fig:arch-0-overview}) consists of a {\it atom-centered descriptor}, a {\it bond-centered descriptor}, a {\it residual dense block}, a {\it readout block} bringing the output irreps to the correct shape, and finally, {\it scale-shift blocks} allowing the even-parity scalars in the readout outputs to be approximately zero-centered, unit-standard-deviation distributions.

We briefly outline the choices in the model.
\begin{enumerate}
    \item The 2B basis is an expansion of Cartesian vectors relative to a central atom in some radial basis function and some subset of spherical harmonics. This a standard approach found in the literature.
    \item The higher-order features are done using a TensorDense operation of the summed 2B features. This is motivated by outer products leading to many-body features as shown in \cite{shapeevMMSMomentTensorPotentials2016,drautzPRBAtomicClusterExpansion2019}. Instead of the full outer product and corresponding contractions, we choose to use a \texttt{TensorDense} layer available in \texttt{e3x}, which linearly projects the incoming 2B features before taking a featurewise tensor product.
    \item The nonlinear block after the descriptor generation and (after optional self-attention message passing) empirically provides improved learnability.
    \item The addition of atom-centered features is motivated by the requirement to have a permutation-symmetric function of two atoms $i$ and $j$.
    \item The bond basis expansion is motivated by the need to incorporate bond orientation data into the learning input. It is performed by using a feature from atom $i$ to atom $j$.
    \item The common residual dense layer is motivated by trying to use the large amount of off-diagonal data to regularize the on-diagonal block learning, as well as to generally add learnable freedom to the model.
    \item The \texttt{TensorDense} readout was chosen to bring the output $\ell$ to the required number for the final prediction head.
\end{enumerate}

In the following sections, we will provide a detailed explanation of the architecture modules, the motivations behind them, and block diagrams detailing their inner workings.

\subsection{Atom-centered descriptor}

\begin{figure}[h]
    \centering
    \includegraphics[width=0.6\linewidth]{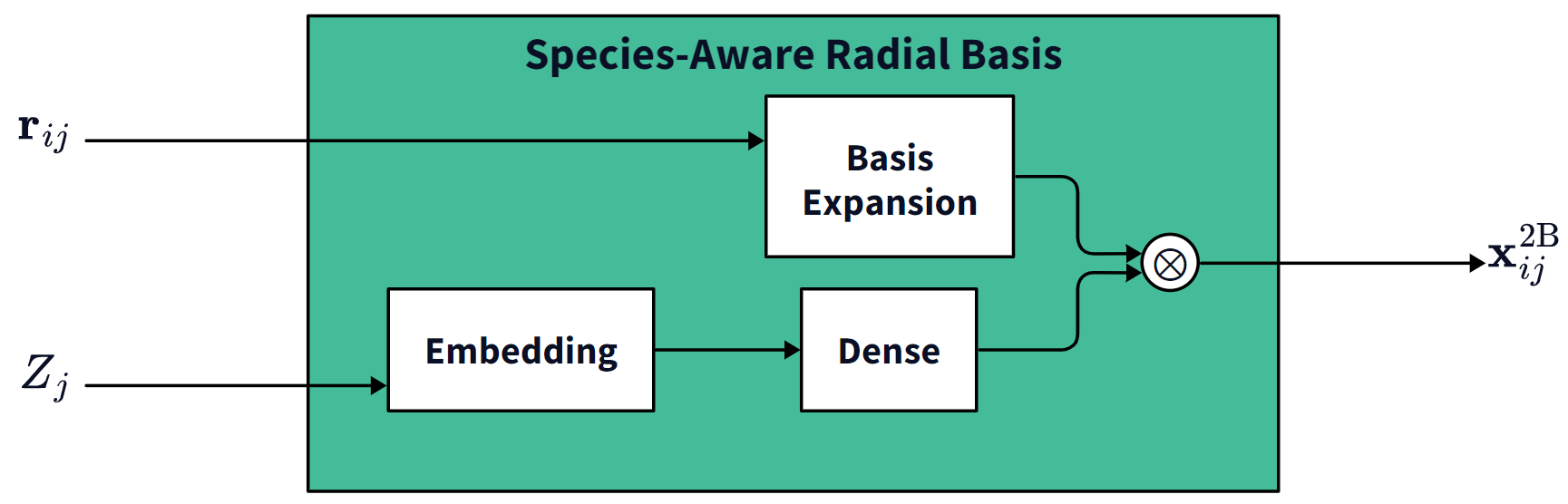}
    \caption{Species aware radial basis block diagram. This layer outputs the 2-body descriptor that is the starting point for the atom-centered block.}
    \label{fig:arch-1-sarb}
\end{figure}

Atom-centered descriptors have been extensively studied in the context of MLIPs; we refer the reader to the review by Musil {\it et al.} \cite{musilCRPhysicsInspiredStructuralRepresentations2021} on their design choices. Here, we illustrate the atom-centered descriptor in GEARS H, which is motivated by previous work in this area.


The first component of the atom-centered descriptor is the species-aware radial Basis, which is diagrammed in Fig. \ref{fig:arch-1-sarb}.
We expand local neighborhoods of atoms using a 2-body (2B) basis consisting of radial and angular functions.
We find that a set of sinusoidal functions with an analytic cutoff to work well.
From previous reports \cite{kocerJCPNovelApproachDescribe2019}, we expect basis functions that converge quickly to a delta function for a suitable linear combination to likely perform better, leading to lower achievable losses.
The $\ell=0_+$ elements of the basis expansion are then multiplied by the species embedding, reshaped to have the correct number of features by a \texttt{Dense} layer.
These are the 2-body (2B) features that the atom-centered block starts with.

\begin{figure}[h]
    \centering
    \includegraphics[width=1.0\linewidth]{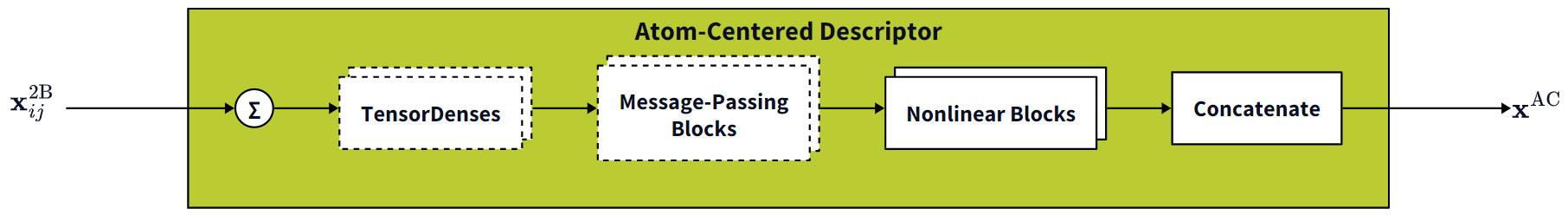}
    \caption{atom-centered descriptor block diagram. This layer outputs the many-body atom-centered descriptors that are the starting point for the bond-centered block and that are used to infer the on-diagonal Hamiltonian irreps.}
    \label{fig:arch-2a-ac}
\end{figure}

\begin{figure}[h]
    \centering
    \includegraphics[width=1.0\linewidth]{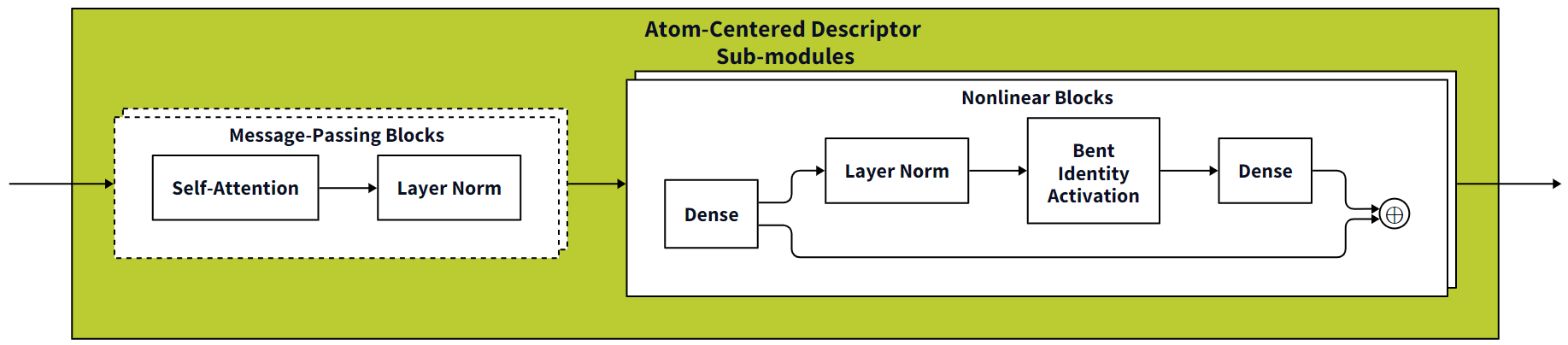}
    \caption{Sub modules of the atom-centered block. These include the Message-Passing blocks and Nonlinear Blocks.}
    \label{fig:arch-2b-ac}
\end{figure}

The atom-centered block is shown in Fig. \ref{fig:arch-2a-ac}, and its submodules are detailed in Fig. \ref{fig:arch-2b-ac}.
The outer product of {\it pooled} 2B features leads to 3-body (3B) features, and subsequent outer products leads to higher-body order features\footnote{This is called the density trick} \cite{shapeevMMSMomentTensorPotentials2016, drautzPRBAtomicClusterExpansion2019, musilCRPhysicsInspiredStructuralRepresentations2021}: here, the pooling operation is over the atoms $j$ in the neighborhood of a given atom $i$. 
The descriptors created from such outer products are known as atom-centered density correlations (ACDC)\cite{nigamJCPUnifiedTheoryAtomcentered2022}.
The 2B features are summed up for all atoms $j$ around atoms $i$ using an \verb|indexed_sum| operation as implemented in E3x.
Empirically, a body order of 3-5 is sufficient for acceptable accuracy of learned quantities like energies and forces.

Motivated by the established body of work on ACDCs, we use a \verb|TensorDense| layer as implemented in E3x \cite{unkeE3xMathrmEquivariant2024} to learn a linear projection of the outer product of 2B features. 
A \verb|TensorDense| layer takes two linear projections of the 2B features followed by a feature-wise tensor product. This corresponds to a linear subspacing operation of a full outer product of the 2B basis. We focus on learning a dense subspace since higher-order descriptors in the Atomic Cluster Expansion (ACE) model are known to be relatively sparse\cite{darbynCMCompressingLocalAtomic2022}.

The 2B descriptor is the radial neighbor density around the $i^{\rm th}$ atom ($\ket{\rho_i}$) as the summation of the neighbor list within the cutoff radius. Similarly, 3B ACDCs can be obtained as the tensor product of 2B ACDCs\cite{nigamJCPEquivariantRepresentationsMolecular2022}.
Therefore, ($N+1$)-body features can be gathered by $N$ times the tensor product of the radial neighbor density.
However, since $\nu^{\rm th}\ $ features, which are $\nu^{\rm th}$ body ACDCs, can be obtained by the tensor product of $(\nu-1)^{\rm th}$features, one can introduce 
\begin{align}
    \ket{\rho_i^{\otimes 2\nu-1}}_< = \mathrm{TensorDense}(\ket{\rho_i^{\otimes \nu}})
    \label{eq:manybody-neighbour-density}
\end{align}

where the subscript $<$ denotes projection to a lower feature dimension.
2B descriptors are passed through \verb|TensorDense| layers (optionally, although we recommend at least one--other wise, the descriptor does not have many-body information) to get 3B descriptors, and so on. 


An essential difference between the well-explored previous energy predictions and our approach is that our model {\it does not} average over all rotations of these ACDCs. In addition, to capture equivariant features from Hamiltonian blocks, our outputs are irreps of Hamiltonian blocks consisting of spherical harmonics of order $\geq 0$.

These 2B, 3B, ..., $\mathrm 2N - 1$B descriptors are then separately (and optionally) message-passed between atoms using self-attention (SA) to carry out the learnable coupling across all incoming messages (see Message-passing block in Fig. \ref{fig:arch-2a-ac}).
To introduce SA, it is important to consider two key facts. First, the query-key (QK) matrices couple messages from atoms $j$ and $j'$, resulting in an additional increase of body order.
Second, the softmax-weighted pooling of the QK matrices leads to coupling across a linear projection of \textit{all} incoming messages.
While normalizing the weights significantly alleviates the need to normalize pooled messages further, it encounters challenges such as outlier features in the QK parameters \cite{heUnderstandingMinimisingOutlier2024}, entropy collapse in the attention blocks \cite{zhaiStabilizingTransformerTraining2023} as well as rank collapse \cite{nociSignalPropagationTransformers2022}. We have not taken any special care against these pitfalls as we have not encountered them, but one must keep them in mind when debugging any potential model performance issues in the future.
After the self-attention step, we have an irrep-wise \verb|LayerNorm| to improve training stability.
The above architectural decisions can be expressed in the following equations (modulo the presence of the \verb|LayerNorm|):

\begin{align}
    \ket{\rho_i^{\otimes [\nu \leftarrow \nu_j]}} = \sum_j \ket{\{\rho_i^{\otimes \nu}\}_i, ...)}
\end{align}

The separate (optionally) message-passed descriptors are then sent through a non-linear block, for which we use a shallow (typically 2-layer) multi-layer perceptron (MLP) with a residual connection, interleaved \verb|LayerNorm|s, and \verb|bent_identity| activation functions to refine the atom-centered features and add more functional expressivity to our descriptor (Fig. \ref{fig:arch-2b-ac}, non-linear block). The user can optimize the MLP architecture and corresponding hyperparameters; our reasonable default setting and details are presented in the following section.

Finally, the resulting descriptors are reduced to a user-controlled maxmimum angular momentum and then concatenated along the feature dimension ($F$ in E3x convention).

By keeping the descriptors of distinct body-order separate until the very end, the intervening message-passing and non-linear blocks remain small (block diagonal in body order), helping reduce parameter counts and speed up training.

\subsection{Bond-centered descriptor}

\begin{figure}[h]
    \centering
    \includegraphics[width=1.0\linewidth]{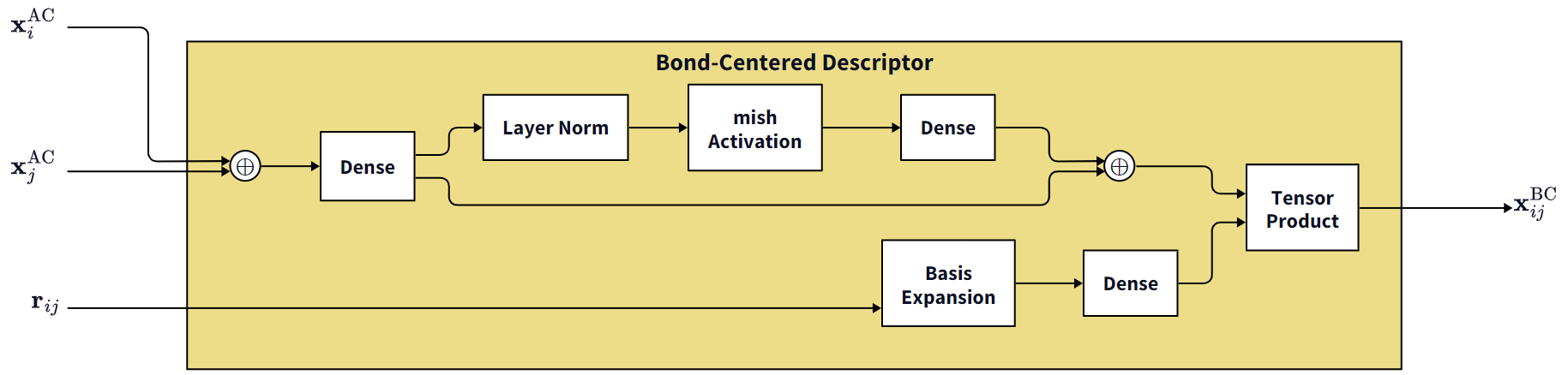}
    \caption{Bond-centered descriptor block. This block combines pairs of atom-centered descriptors to make the bond-centered descriptors, which are then used to infer the off-diagonal Hamiltonian irreps.}
    \label{fig:arch-3-bc}
\end{figure}

The bond-centered descriptor is shown in Fig. \ref{fig:arch-3-bc}.
Off-diagonal terms in Hamiltonian or overlap matrix blocks can be predicted as a function of atom-centered features of the two atoms comprising a `bond'. A bond here concretely refers to any two atoms with significant basis function overlap (and corresponding interaction strength), which differs from the atom-in-molecule definition using bond critical points of the electron density \cite{baderAtomsMoleculesQuantum2003}.
To calculate atom-pairwise features for predicting off-diagonal matrix blocks, we take a summation of atom-centered features, a permutation-invariant pooling operator, which is a similar approach with PhiSNet \cite{unkeSE3equivariantPredictionMolecular2021}. 
To add more functional expressivity, the pooled features are refined using a shallow MLP with \verb|LayerNorm|, \verb|mish| activation, and a residual connection as in the atom-centered case.
For bond orientation information, we expand the bond vector on a basis function and have a \verb|Dense| layer as a learnable linear projection to refine features. 
We tensor product a linear projection of the bond vector feature-wise to the pooled atom-pair features. This final tensor product incorporates bond orientation information into the pooled atom-centered features.

\subsection{Residual dense block}

\begin{figure}[h]
    \centering
    \includegraphics[width=1.0\linewidth]{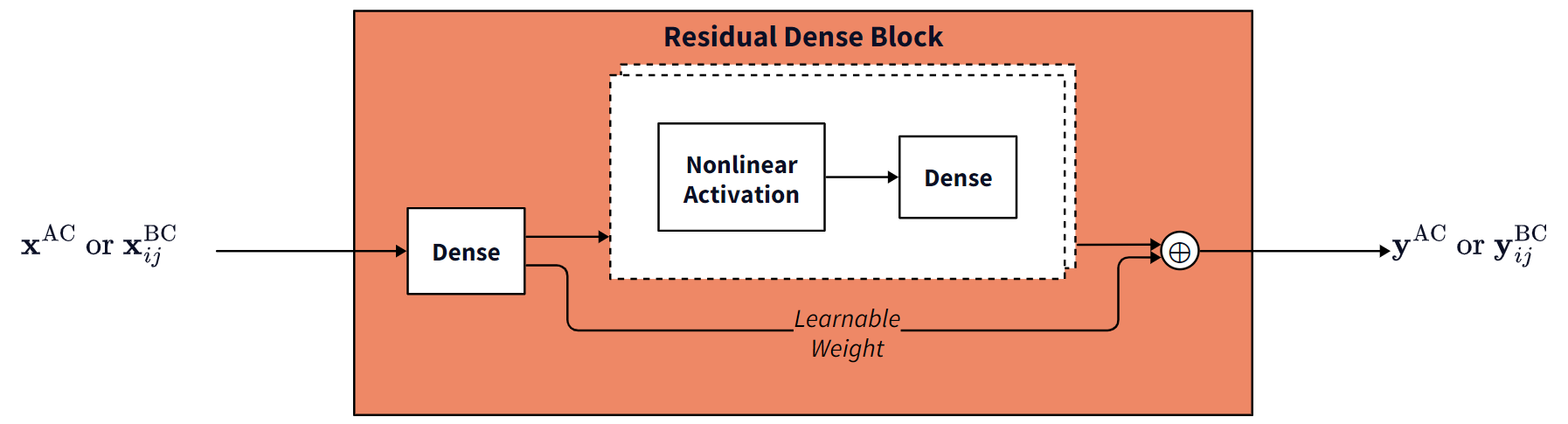}
    \caption{Residual Dense block. atom-centered descriptors and bond-centered descriptors are pass through the same block before being sent to their respective readouts and Scale Shift blocks.}
    \label{fig:arch-4-rd}
\end{figure}

We use a residual-type dense network to refine the final orientation-aware atom-pair features (for off-diagonal Hamiltonian blocks) or atom-centered features (for on-diagonal Hamiltonian blocks).
The number and size of the dense layers and nonlinear activation function used are user controllable.


\subsection{Readout blocks}
The outputs of the residual dense block are passed through two separate \verb|Readout| blocks, for off-diagonal and on-diagonal readouts. These are \verb|TensorDense| blocks that reshape the output of the residual dense blocks to the correct feature count and angular momentum to correspond to the irreps of the Hamiltonians being predicted.

\subsection{Scale-shift blocks}
This is a non-learnable block that scales and shifts the parity-symmetric scalars in the readout output. This leads to (scalar) outputs from the readout that can be approximately-zero-centered, approximately-unit-variance and map them to numbers which may be very far from 0.

Specifically, the orbital-diagonal (equal angular momentum for both orbitals) blocks of the on-diagonal $H$ blocks contain parity-symmetric scalars that can get quite large in magnitude. More generally, any on-diagonal $H$ block between two equal angular momentum orbitals will have a parity-symmetric scalar due to the Wigner-Eckhart theorem (since the lower limit of $|\ell_1 - \ell_2|$ is now $0$, and all equal-angular-momentum blocks are even parity).
Empirically, we have seen that for semicore orbitals or polarization orbitals, these scalars can reach $\geq125$ in magnitude and span a range of 200 across features.
More importantly, for a given block, the actual \textit{environment-dependent spread} around the typical mean value is small compared to the mean value itself.
This presents an ideal use case for scale-shifting the scalars.

For off-diagonal blocks, we notice a similar pattern: Often, a part of the variation of parity-symmetric scalars can be expressed purely as a function of the distance of two atoms.  Motivated by this observation, we implement a shift operation for the parity-symmetric scalars of off-diagonal $H$ irreps that is dependent purely on element-pairs and radial separation between the atoms.

With these architectural choices, the Hamiltonian blocks corresponding to pairs of atoms are approximated as E(3)-equivariant functions of local atomic environments with species-permutation symmetry. We have not yet built in the Hamiltonian symmetry (transpose of outer product) into our readouts.
This is added as a post-processing step where we define our final Hamiltonian as:

\begin{align}
    H_{final} = \frac{1}{2} (H_{predicted} + H_{predicted}^T)
\end{align}

\subsection{Additional architecture features}


We use a learnable normalization after the self-attention steps and after the bond-pooling steps. Empirically that this leads to more robust training. There is some evidence that layer normalization leads to outlier features \cite{heUnderstandingMinimisingOutlier2024} but this only affects low-precision pooling operations, which is not the case for us as we train and infer in 32-bit. Nevertheless, this can be a promising avenue of further investigation.



We use a combination of mean-squared error (MSE) and root-mean-squared error (RMSE) as the default loss function, with the weights adjustable by the user.
This loss is invariant to rotations of vectors, whereas the mean absolute error is not \cite{wangDesignSpaceMolecular2024}. RMSE loss functions have a nonzero gradient at the first order, even in the neighborhood of 0 difference between ground truth data and predicted output.




\section{Training hyperparameters}
\label{s:S-hyperparams}

Each of the following sections include a portion of the full configuration file. 
There is very little variation in the model hyperparameters used to train the models presented in this work.
This highlights the robustness of the default parameters used in GEARS H and enables the research community to apply GEARS H with ease.
As such, we will only discuss what changes were made relative to each section of the configuration in the subsequent sections.

\subsection{Data}

\begin{listing}[h]
\begin{minted}
[
frame=lines,
framesep=2mm,
baselinestretch=1.2,
fontsize=\scriptsize,
bgcolor=backcolour,
linenos
]
{yaml}
data:
  directory: <omitted>
  experiment: <omitted>
  train_data_path:
  - <omitted>
  val_data_path:
  - <omitted>
  n_train: 20
  n_valid: 20
  bond_fraction: 0.3
  sampling_alpha: 0.0
  atoms_pad_multiple: 100
  nl_pad_multiple: 10000
  batch_size: 1
  valid_batch_size: 1
  shuffle_buffer_size: 100
  energy_unit: eV
  pos_unit: Ang
\end{minted}
\caption{Hyperparameters related to the training and validation data.}
\label{listing:data}
\end{listing}

Hyperparameters related to the training and validation data are shown in Listing \ref{listing:data}. 
The root directory used, the experiment name, and actual training and validation dataset paths are omitted for brevity.
\verb|n_train|, \verb|n_valid|, \verb|atoms_pad_multiple|, and \verb|nl_pad_multiple| are changed depending on the dataset.
The first two control the number of training and validation structures, while the last two control the number of recompilations of the model (through the maximum amount of padding a neighbor list array and an atomic species array is permitted) that the user is willing to allow.
More recompilations means faster epochs (due to less padding and carrying of otherwise unnecessary zeroes through the training process) at the expense of a slower start, while fewer compilations enables a quicker start at the expense of slower epochs.

\subsection{Atom-centered descriptor}

\begin{listing}[h]
\begin{minted}
[
frame=lines,
framesep=2mm,
baselinestretch=1.2,
fontsize=\scriptsize,
bgcolor=backcolour,
linenos
]
{yaml}
model:
  atom_centered:
    descriptor:
      descriptor_name: ShallowTDSAAtomCenteredDescriptor
      use_fused_tensor: True
      num_tensordenses: 1
      max_tensordense_degree: 4
      num_tensordense_features: 12
      mp_steps: 0
      mp_degree: 4
      mp_options:
        num_heads: 4
        qkv_features: 32
      mp_basis_options:
        cutoff_fn: smooth_cutoff
        radial_fn: basic_fourier
        radial_kwargs: {}
        max_degree: 2
        num: 8
    radial_basis:
      cutoff: 8.0
      num_radial: 24
      max_degree: 4
      num_elemental_embedding: 32
\end{minted}
\caption{Atom-centered descriptor and radial basis parameters used in training models with 1 \texttt{TensorDense} operation.}
\label{listing:acrb}
\end{listing}

atom-centered and Species-Aware Radial Basis hyperparameters are shown in Listing \ref{listing:acrb}.
The only change in atom-centered hyperparameters across models is the number of \verb|TensorDense|s, which was set to either 1 or 2 for all models shown in this work.
Note that the message-passing options (which all begin with \verb|mp_|) are all unnecessary because \verb|mp_steps| is set to 0 for all models.
We leave them here for completeness.

The only changes made in the Species-Aware Radial Basis hyperparameters were to adjust the cutoff radius, which was adjusted to the maximum basis set cutoff across species in the dataset.
Basis set cutoffs are discussed in Section 4.1.2.

\subsection{Bond-centered descriptor}

\begin{listing}[h]
\begin{minted}
[
frame=lines,
framesep=2mm,
baselinestretch=1.2,
fontsize=\scriptsize,
bgcolor=backcolour,
linenos
]
{yaml}
model:
  bond_centered:
    cutoff: 8.0
    max_basis_degree: 4
    max_degree: 4
    tensor_module: fused_tensor
    tensor_module_dtype: float32
    bond_expansion_options:
      cutoff_fn: smooth_cutoff
      radial_fn: basic_fourier
      radial_kwargs: {}
      max_degree: 4
      num: 24
\end{minted}
\caption{bond-centered descriptor hyperparameters. Note that the \texttt{model:} tag here is redundant with the one shown in Listing \ref{listing:acrb}--only one \texttt{model:} tag should be present in the input file.}
\label{listing:bc}
\end{listing}

The bond-centered hyperparameters are shown in Listing \ref{listing:bc}. These were left unchanged across all models shown in this work, except for the cutoff, which was set the the largest cutoff across species in the dataset (see Section 4.1.2).

\subsection{MLP}

\begin{listing}[h]
\begin{minted}
[
frame=lines,
framesep=2mm,
baselinestretch=1.2,
fontsize=\scriptsize,
bgcolor=backcolour,
linenos
]
{yaml}
model:
  mlp:
    mlp_layer_widths: [32,16,32]
    mlp_dtype: float32
    mlp_activation_function: bent_identity
\end{minted}
\caption{Multilayer perceptron hyperparameters (used in the Residual Dense layer). Note that the \texttt{model:} tag here is redundant with the one shown in Listing \ref{listing:acrb}--only one \texttt{model:} tag should be present in the input file.}
\label{listing:rd}
\end{listing}

The MLP hyperparamters that define the Residual Dense block are shown in Listing \ref{listing:rd}.
These parameters are unchanged across all models shown.

\subsection{Optimizer and learning rate schedule}

\begin{listing}[h]
\begin{minted}
[
frame=lines,
framesep=2mm,
baselinestretch=1.2,
fontsize=\scriptsize,
bgcolor=backcolour,
linenos
]
{yaml}
optimizer:
  lr: 0.005
  name: adan
  opt_kwargs:
    weight_decay: 0.001
  schedule:
    name: reduce_on_plateau
    factor: 0.9
    patience: 50
    min_scale: 0.01
    rtol: 0.01
    atol: 0
    cooldown: 25
    accumulation_size: 160
\end{minted}
\caption{Optimizer and learning rate schedule hyperparameters.}
\label{listing:opt}
\end{listing}

Hyperparameters defining the optimizer and learning rate (LR) schedule used to train the models in this work are shown in Listing \ref{listing:opt}.
The only optimizer changes made were to switch between \texttt{adan} \cite{xieAdanAdaptiveNesterov2024} to \texttt{lamb} \cite{youLargeBatchOptimization2020} as we found one or the other performed better on some datasets.

The only LR schedule changes made were to adjust the \texttt{accumulation\_size} parameter to make the \texttt{reduce\_on\_plateau} scheduler check if the loss had plateaued once per epoch.

\subsection{Loss}

\begin{listing}[h]
\begin{minted}
[
frame=lines,
framesep=2mm,
baselinestretch=1.2,
fontsize=\scriptsize,
bgcolor=backcolour,
linenos
]
{yaml}
loss:
  name: weighted_mse_and_rmse
  loss_parameters:
    off_diagonal_weight: 4.0
    on_diagonal_weight : 1.0
    mse_weight         : 1.0
    rmse_weight        : 1.0
    loss_multiplier    : 5.0
\end{minted}
\caption{Loss calculation hyperparameters used to train all models.}
\label{listing:loss}
\end{listing}

Loss calculation hyperparameters are shown in Listing \ref{listing:loss}. 
These were left unchanged across all models trained.
We weight the off-diagonal loss $4\times$ that of the on-diagonal loss, and multiply the total loss by $5\times$ to ensure we avoid the \texttt{float32} precision floor.

\section{Mapping of Hamiltonian blocks to readout features}

\verb|e3x| has equivariant features in the format of \verb|A_pLF|, where \verb|p| is the parity axis, \verb|L| is the angular momentum index, and \verb|F| is the feature index. More details and intuitive visualizations can be found in \cite{unkeE3xMathrmEquivariant2024}, particularly figure 2 in the reference.
We map the H blocks to the features by looping over the orbitals in the rows and columns of the H block. For each orbital pair, we take the corresponding subblock, then unwrap it from direct product to direct sum form. For example, let us consider an H block between two atoms of different species with L=0,1,2 orbitals ($s$, $p$, $d$). There are the following distinct subblocks: $ss$, $sp$, $sd$, $ps$, $pp$, $pd$, $ds$, $dp$, $dd$. Their direct sum forms, from Wigner-Eckhart theorem, are (we denote even and odd using e and o in the following list)

\begin{itemize}
    \item $ss$: L=0, e
    \item $sp$: L=1, o
    \item $sd$: L=2, e
    \item $ps$: L=1, o
    \item $pp$, L=0+1+2, e
    \item $pd$, L=1+2+3, o
    \item $ds$, L=2, e
    \item $dp$, L=1+2+3, o
    \item $dd$, L=0+1+2+3+4, e
\end{itemize}

The features are chosen to be the first feature which has all relevant angular momenta channels not bound to any prior H sub-block. The concrete implementation details are in \verb|https://github.com/SamsungDS/gears_h/blob/main/gears_h/utilities/mapmaker.py|.

\section{Comments on inference}

The full inference process on the structures used in the device-scale structure section takes approximately \SI{2.3}{\minute} using an AMD 9684X CPU and \SI{1.3}{\minute} using an Nvidia L40S GPU. 
CPU and GPU inference alone took 13 seconds and 19 seconds, respectively. 
The remainder of the time was spent on combining Hamiltonian irreps to get Hamiltonian blocks, and then assembling and writing the final Hamiltonian.

GPUs make extensive use of low-precision operations to accelerate computation. Left unaddressed, this leads to errors during inference. Through \verb|jax|, we enforce the highest precision possible for GPU operations, mitigating the precision errors on the GPUs we tested. However, since we cannot test across all GPU models and other machine-learning specialized devices, out of an abundance of caution, GEARS H infers using the CPU by default. Regardless of the device used for inference, the speedup over a self-consistent solution of the Kohn-Sham equations is immense.

\section{Hamiltonian off- and on-diagonal block errors}

\begin{figure}[h!]
    \centering
    \includegraphics[width=1.0\linewidth]{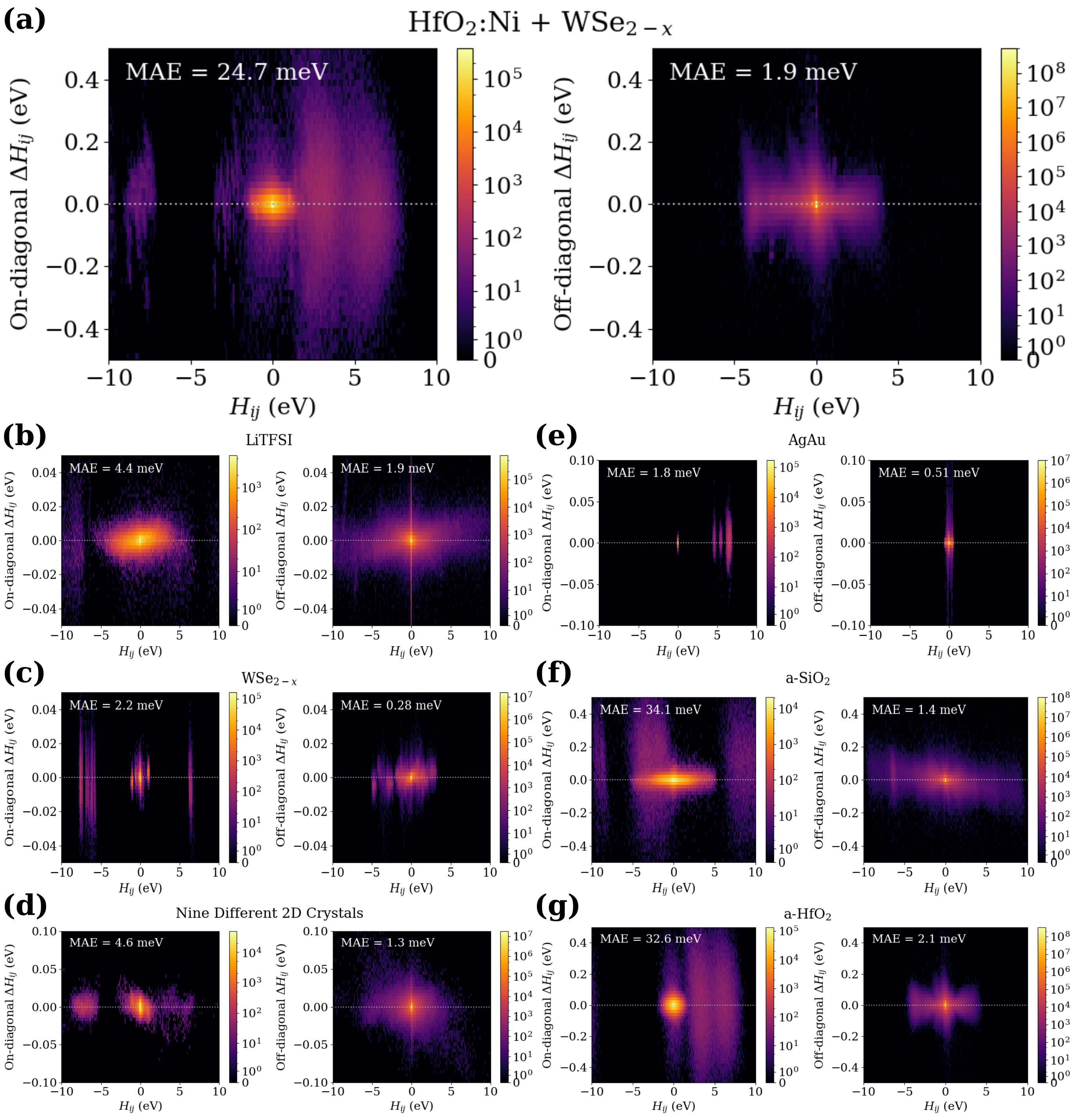}
    \caption{Hamiltonian matrix element errors split into on-diagonal and off-diagonal components for all models presented in the main body.}
    \label{sfig:onofferrors}
\end{figure}

In Fig. \ref{sfig:onofferrors}, we show the Hamiltonian matrix element errors for every model show in the paper divided into on-diagonal block errors (left subplots) and off-diagonal block errors (right subplots).
In all models, the on-diagonal block MAE is larger, often by a full order of magnitude.
This is slightly counter-balanced by the fact that the on-diagonal matrix elements are larger than the off-diagonal matrix elements.
Nevertheless reducing these errors presents an important target for improvement in future versions of GEARS H.

\section{Effect of adding additional message-passing steps}

\begin{figure}[h!]
    \centering
    \includegraphics[width=0.75\linewidth]{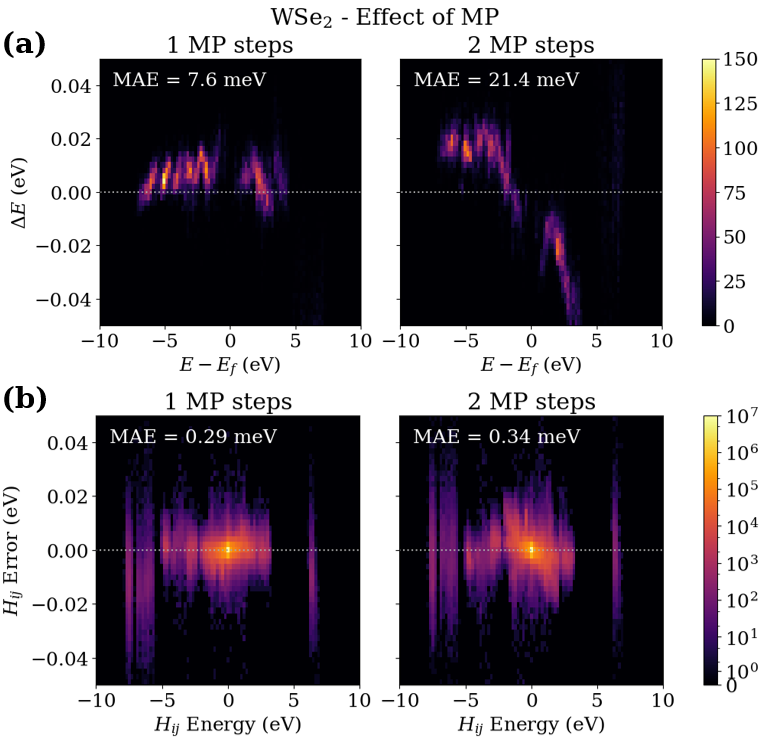}
    \caption{Errors of $\ce{WSe_{2-x}}$ models that include 1 and 2 message-passing steps. (a) Eigenvalue errors. (b) Hamiltonian matrix element errors.}
    \label{sfig:wse2_mp_h_errors}
\end{figure}

In Fig. \ref{sfig:wse2_mp_h_errors}, we show the eigenvalue errors and Hamiltonian matrix element errors for $\ce{WSe_{2-x}}$ models trained with one and two message-passing steps.
Besides the addition of the message-passing steps and their hyperparameters, all hyperparameters are identical to those used for training the $\ce{WSe_{2-x}}$ model shown in the main body.

We use 4 heads and 32 features for the queries, keys, and values.
The maximum degree in the message-passing blocks is set to $\ell = 4$, and the message-passing basis expansion uses 8 radial functions with a maximum degree of 2.
The model with no message-passing has 82,909 parameters.
Adding one message-passing step increases the parameter count to 125,545 parameters, and adding two message-passing steps increases the parameter count to 180,981 parameters.

The MAE of the Hamiltonian matrix elements of these models are comparable to those of the model with no message-passing steps, while the eigenvalue errors are larger.
Despite the significant increase in parameter count, or perhaps because of, the models are not better.
In our experience, the addition of message-passing steps for more complex systems is even worse than in this example.

\section{Effect of training set size}

\begin{figure}[h!]
    \centering
    \includegraphics[width=0.75\linewidth]{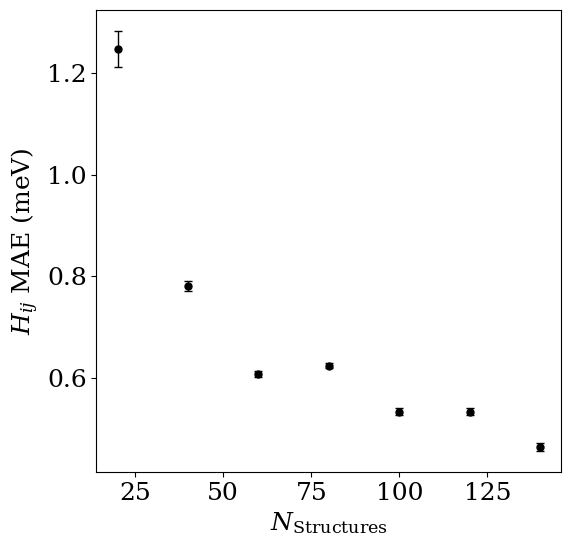}
    \caption{Hamiltonian matrix element errors as a function of training set size for $\ce{WSe_{2-x}}$ models.}
    \label{sfig:wse2_training_set_size_hij_errors}
\end{figure}

In Fig. \ref{sfig:wse2_training_set_size_hij_errors}, we show the Hamiltonian matrix element MAE as a function of training set size (ranging from 20 to 140 structures) for $\ce{WSe_{2-x}}$ models with the exact same model hyperparameters as the one shown in the main text.
The decrease in the MAE is rapid with the addition of new training structures for small training sets, but begins to flatten as more structures are added.

\section{Inference on crystalline \ce{SiO2} polymorphs}

\begin{figure}[h!]
    \centering
    \includegraphics[width=0.75\linewidth]{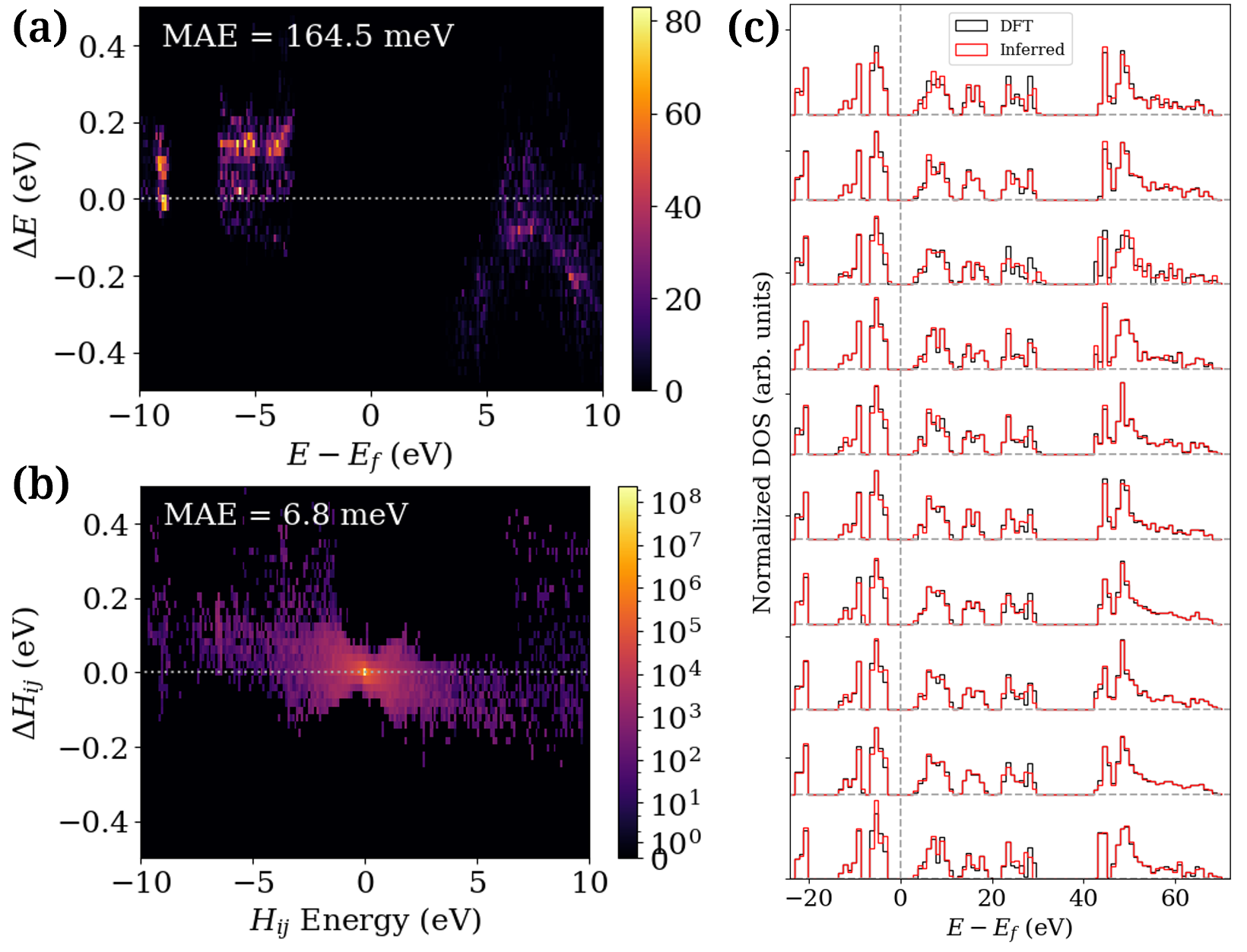}
    \caption{Hamiltonian matrix element errors as a function of training set size for $\ce{WSe_{2-x}}$ models.}
    \label{sfig:crystalline_sio2}
\end{figure}

In Fig. \ref{sfig:crystalline_sio2}, we show the eigenvalue errors, Hamiltonian matrix element errors, and histograms of the inferred and reference eigenvalue spectrum for inference on 10 different \ce{SiO2} polymorphs inferred using the a-\ce{SiO2} model.
In order from bottom to top of Fig. \ref{sfig:crystalline_sio2}(c), the Materials Project IDs for these polymorphs are  556961, 640556, 542814, 733790, 554573, 555235, 8059, 559091, 554089, and 546794. 
The density range of the amorphous model training set spans the range of these crystalline polymorphs, leading to acceptably good predictions.
Accuracy can be improved by including near-crystalline structures in the training set and/or increasing the training set size.
This is a simple demonstration of the model's ability to perform far out of domain.

\clearpage

\printbibliography

\end{document}